\documentclass[iop]{emulateapj}
\usepackage{lscape}
\usepackage{rotating}
  \received{}
  \revised{}
  \accepted{}

\shortauthors{Auger et al.}
\shorttitle{Colors, Lensing and Stellar Masses of Early-type Galaxies}

\begin{document}

\title{The Sloan Lens ACS Survey. IX. Colors, Lensing and Stellar Masses of Early-type Galaxies}

\author{M. W. Auger\altaffilmark{1,$\dagger$}, T. Treu\altaffilmark{1,2}, A. S. Bolton\altaffilmark{3}, R. Gavazzi\altaffilmark{4}, L. V. E. Koopmans\altaffilmark{5}, P. J. Marshall\altaffilmark{1}, K. Bundy\altaffilmark{6}, L. A. Moustakas\altaffilmark{7}}
\altaffiltext{1}{Department of Physics, University of California, Santa Barbara, CA 93106, USA}
\altaffiltext{2}{Sloan Fellow, Packard Fellow}
\altaffiltext{3}{Beatrice Watson Parrent Fellow, Institute for Astronomy, University of Hawai'i, 2680 Woodlawn Dr., Honolulu, HI 96822}
\altaffiltext{4}{Institut d'Astrophysique de Paris, UMR7095 CNRS \& Univ. Pierre et Marie Curie, 98bis Bvd Arago, F-75014 Paris, France}
\altaffiltext{5}{Kapteyn Astronomical Institute, University of Groningen, P.O. Box 800, 9700AV Groningen, The Netherlands}
\altaffiltext{6}{Hubble Fellow, Department of Astronomy, University of California, Berkeley, CA 94720}
\altaffiltext{7}{Jet Propulsion Laboratory, California Institute of Technology, 4800 Oak Grove Drive, M/S 169-237, Pasadena, CA 91109}
\altaffiltext{$\dagger$} {\texttt mauger@physics.ucsb.edu}

\begin{abstract}
We present the current photometric dataset for the Sloan Lens ACS
(SLACS) Survey, including HST photometry from ACS, WFPC2, and
NICMOS. These data have enabled the confirmation of an additional 15
grade `A' (certain) lens systems, bringing the number of SLACS grade
`A' lenses to 85; including 13 grade `B' (likely) systems, SLACS has
identified nearly 100 lenses and lens candidates. Approximately 80\%
of the grade `A' systems have elliptical morphologies while $\sim $10\% show
spiral structure; the remaining lenses have lenticular morphologies.
Spectroscopic redshifts for the lens and source are available for
every system, making SLACS the largest homogeneous dataset of
galaxy-scale lenses to date. We have created lens models using
singular isothermal ellipsoid mass distributions for the 11 new
systems that are dominated by a single mass component and where
the multiple images are detected with sufficient signal-to-noise; these
models give a high precision measurement of the mass within the
Einstein radius of each lens. We have developed a novel
Bayesian stellar population analysis code to determine robust stellar
masses with accurate error estimates. We apply this code to deep,
high-resolution HST imaging and determine stellar masses with typical
statistical errors of 0.1 dex; we find that these stellar masses
are unbiased compared to estimates obtained using SDSS photometry, provided
 that informative priors are used. The stellar masses range from $10^{10.5}$
to $10^{11.8}$ M$_\odot$ and the typical stellar mass fraction within
the Einstein radius is 0.4, assuming a Chabrier IMF. The ensemble
properties of the SLACS lens galaxies, e.g. stellar masses and projected
ellipticities, appear to be indistinguishable from other SDSS galaxies
with similar stellar velocity dispersions. This further supports
 that SLACS lenses are representative of the overall population
of massive early-type galaxies with M$_* \gtrsim 10^{11} {\rm M_\odot}$,
and are therefore an ideal dataset to investigate the kpc-scale
distribution of luminous and dark matter in galaxies out to $z \sim 0.5$.
\end{abstract}

\keywords{
   galaxies: elliptical and lenticular, cD -- gravitational lensing -- surveys
}

\section{INTRODUCTION}
Early-type galaxies have been found to be remarkably homogeneous, as
evidenced by the tight scaling relationships of the Fundamental Plane
\citep{fp,dressler} and the small scatter of their nearly-isothermal
central density profiles \citep[e.g.,][]{rk,rkk,paper3,koopmans09}. However,
there are many details of this homogeneity that remain unresolved, including
an explanation for the tilt of the Fundamental Plane with respect to the
virial plane and for the scatter, or finite thickness, orthogonal to the
plane. \citet{paper7} have used lensing and stellar velocity
dispersions to show that the dynamical mass scales with the lensing
(i.e., total) mass of early-type galaxies, suggesting that non-homology
is unlikely to cause the observed tilt. Joint dynamical and stellar
population modeling has suggested that more massive early-type galaxies have
higher dark matter fractions \citep[e.g.,][]{padmanabhan} that could
explain most of the tilt \citep{tortora}. These models depend on
assumptions about the initial mass function and do not readily
explain how the luminous bulge and a dark matter halo might `conspire'
to produce an isothermal total mass density profile; a well-defined
set of galaxies with accurate photometric data is required in order
to perform the detailed modeling necessary to accurately separate
the dark and luminous mass components.

A particularly powerful tool is the
combination of precise strong gravitational lensing mass measurements,
stellar velocity dispersions, and accurate stellar mass estimates.
The advent of the Sloan Lens ACS Survey \citep[SLACS;][papers I-VIII,
respectively]{paper1,paper2,paper3,paper4,paper5,paper6,paper7,paper8}
has made this type of modeling feasible for the first time for a large
and uniformly selected dataset. One of the key features of SLACS is
that each lens contains all of the information necessary to perform a
robust decomposition of the matter content in the central regions of
galaxies. For example, the SDSS spectroscopy provides a precise
measurement of the stellar velocity dispersion and the lens and source
redshifts, while high-resolution Hubble Space Telescope (HST) imaging
yields a detailed view of the lensed background source and the surface brightness profile of the lensing galaxy.

In this paper we present the definitive SLACS sample and an essential
component necessary for disentangling the dark and luminous matter in
early-type galaxies: multi-band optical and near-infrared imaging from
HST. The photometric colors provided by HST allow us to accurately
constrain the stellar mass of the lensing galaxy, thereby determining
the normalization of the luminous component of the mass
distribution. Although other authors have investigated the stellar
populations of SLACS lenses using SDSS photometry \citep[][hereafter
G09]{grillo,grillo09}, we find that the HST imaging is necessary to
adequately perform model photometry and yield unbiased stellar mass
estimates, to the level of accuracy that is comparable with that of
the other observables.

The paper is organized as follows.  We begin by describing our new HST
observations in Section 2. We then provide a detailed summary of the
SLACS dataset in Section 3, including the introduction of 15 newly
confirmed strong lenses. In Section 4 we discuss our model photometry,
and our stellar mass estimates are detailed in Section 5. A discussion
of the ensemble properties of SLACS lenses, including comparison with
`twin' galaxies from the SDSS, is given in Section 6 before concluding
with a summary of the final SLACS dataset in Section 7. All magnitudes
are on the AB scale \citep{oke}, a standard concordance cosmology with
$\Omega_{m}=0.3$ $\Omega_{\Lambda}=0.7$ $h=0.7$ is assumed, and base-10
logarithms are denoted by `log' while natural logs are written as `ln.'

\section{Multi-Color Imaging}
A campaign was initiated to obtain HST multi-band imaging in V, I, and
H (F555W, F814W, and F160W) for all of the confirmed SLACS lenses and
a small number of high-probability lens candidates from Paper V using the
ACS WFC and NICMOS NIC2 cameras (Programs 10494, 10798, and 11202; PI
Koopmans). The programs were switched to WFPC2 after the failure of
ACS, and F555W was replaced with F606W to exploit the improved
throughput in F606W compared to F555W on WFPC2; we refer to these two
filters interchangeably as the V-band throughout the rest of this paper.
Furthermore, the failure of NICMOS in September 2008 has left 29 lenses
unobserved in the H band. All of the ACS data were reduced using the methods
described in Paper V. A total of 77 systems were observed with WFPC2 and 59
systems were observed with NICMOS; this includes 12 new lens candidates not
included in Paper V (Table \ref{T_new_targets}). A summary of the
imaging for our targets is given in Table \ref{T_filter_info}. 

\begin{deluxetable*}{lrrlllc}
\tabletypesize{\scriptsize}
\tablecolumns{7}
\tablewidth{0pc}
\tablecaption{New SLACS Targets}
\tablehead{
 \colhead{Name} &
 \colhead{Right Ascension} &
 \colhead{Declination} &
 \colhead{$z_{fg}$}  &
 \colhead{$z_{bg}$}  &
 \colhead{$\sigma_{\rm SDSS}$ (km s$^{-1}$)} &
 \colhead{Grade}
}
\startdata
SDSSJ1038$-$0024  &  10h38m58.45s  &  -00$^{\circ}$24m38.6s  &  0.219  &  1.306  &  198$\pm$11  & B \\
SDSSJ1306$+$0600  &  13h06m13.65s  &  06$^{\circ}$00m22.1s  &  0.173  &  0.472  &  237$\pm$16  & A \\
SDSSJ1313$+$4615  &  13h13m02.93s  &  46$^{\circ}$15m13.6s  &  0.185  &  0.514  &  262$\pm$18  & A \\
SDSSJ1318$-$0313  &  13h18m39.33s  &  -03$^{\circ}$13m34.2s  &  0.240  &  1.300  &  213$\pm$17  & A \\
SDSSJ1319$+$1504  &  13h19m32.50s  &  15$^{\circ}$04m26.0s  &  0.154  &  0.606  &  235$\pm$13  & A \\
SDSSJ1325$+$4145  &  13h25m55.31s  &  41$^{\circ}$45m02.2s  &  0.306  &  0.785  &  259$\pm$15  & B \\
SDSSJ1436$+$3640  &  14h36m05.63s  &  36$^{\circ}$40m47.1s  &  0.185  &  0.758  &  280$\pm$18  & A \\
SDSSJ1527$+$0314  &  15h27m47.69s  &  03$^{\circ}$14m32.1s  &  0.222  &  0.538  &  231$\pm$16  & X \\
SDSSJ1644$+$2625  &  16h44m43.09s  &  26$^{\circ}$25m25.4s  &  0.137  &  0.610  &  229$\pm$11  & A \\
SDSSJ1709$+$2324  &  17h09m38.96s  &  23$^{\circ}$24m08.4s  &  0.347  &  0.719  &  286$\pm$30  & A \\
SDSSJ1719$+$2939  &  17h19m34.16s  &  29$^{\circ}$39m26.4s  &  0.181  &  0.578  &  286$\pm$15  & A \\
SDSSJ2343$-$0030  &  23h43m58.87s  &  -00$^{\circ}$30m22.4s  &  0.181  &  0.463  &  268$\pm$10  & A \\
\vspace{-7pt}
\enddata
\label{T_new_targets}
\end{deluxetable*}

\begin{deluxetable}{lccc}
 \tabletypesize{\scriptsize}
 \tablecolumns{4}
 \tablewidth{0pc}
 \tablecaption{SLACS Imaging}
 \tablehead{
  \colhead{Photometry} &
  \colhead{\# A Lenses} &
  \colhead{\# B Lenses} &
  \colhead{\# X Targets}
 }
\startdata
       F435W  &  33  &   6  &  18  \\
       F555W  &  16  &   0  &   0  \\
       F606W  &  64  &   8  &   5  \\
       F814W  &  82  &  11  &  44  \\
       F160W  &  51  &   3  &   1  \\
    One band  &   6  &   4  &  24  \\
   Two bands  &  22  &   3  &  19  \\
 Three bands  &  32  &   6  &   2  \\
  Four bands  &  25  &   0  &   0  \\
\vspace{-7pt}
\enddata
 \label{T_filter_info}
\end{deluxetable}

\subsection{WFPC2 Data Reduction}
The SLACS WFPC2 images were reduced using a custom-built pipeline that largely follows the procedures used in the PyRAF task {\sc multidrizzle}. One significant difference is the initial step; we run a Python implementation of the {\sc lacosmic} program on each of the raw science frames to identify and mask cosmic rays. The individual exposures are then background subtracted and the {\sc crossdriz} task is used to find the sub-pixel shifts between exposures. We only use the shifts determined from the WFC3 chip because we know this chip contains a bright source (the SLACS target) to ensure robust cross correlation. This avoids problems with spurious offsets from the other chips (due to no objects in the chip frame, for example) skewing the average offset; this procedure yields image registration to better than 0.1 pixel. The 4 exposures for each chip are drizzled to an output image using a 0\farcs05 pixel scale and a pixfrac of 0.6. The drizzled images of the four chips are then placed into a common image using SWARP. All of the lens-galaxy photometry presented in this paper is performed on the pre-SWARP drizzled image of the WFC3 chip. Absolute astrometry is achieved by setting the location of the lens galaxy to the SDSS coordinates.

\subsection{NICMOS Data Reduction}
The SLACS lens candidates tend to be quite large compared to the NICMOS field of view, and as a result special care must be taken when reducing the NICMOS imaging. We have used the most recent version of the CALNICA pre-processing pipeline (version 4.4.1, implemented in the OPUS pipeline and applied by re-downloading the \_cal.fits images from the HST archive) to perform the flatfielding, de-biasing, linearity corrections, time-stream cosmic ray rejection, bad-pixel masking, and error model creation.

One particular problem presented by the large size of the SLACS galaxies is that the unknown floating bias of each amplifier (the `pedestal') cannot be accurately quantified using standard techniques if the lens dominates the amplifier quadrant. We use an iterative approach to determine an approximation of the true pedestal by first using the STSDAS task {\sc pedsub} to remove an initial estimate of the pedestal. We next fit a model of the amplifier glow in the pedestal-subtracted images to remove residuals which were not adequately subtracted by CALNICA. The scaled amplifier glow model is then subtracted from the original data and {\sc pedsub} is performed again on the improved amplifier glow-corrected images. At this point most of the instrumental signature has been removed from the data and we apply the count-rate non-linearity correction implemented in the Python task {\sc rnlincor}. A robust clipping/fitting algorithm is then employed to determine the sky level in each exposure; this sky is subtracted and the STSDAS {\sc crossdriz} routine is used to determine offsets between exposures, yielding registration to better than 0.1 pixel. The exposures are then drizzled into a median image using the STSDAS {\sc drizzle} task. As in the standard reduction pipelines, the median image is blotted back to the original CCD frame for residual cosmic ray/bad pixel detection. However, we also use the blotted median image to subtract a model of the astrophysical sources, leaving residuals which are dominated by the quadrant-dependent errors in the pedestal removal. The residual bias in each quadrant is found by an aggressively clipped mean and removed. The median image used as a model for the astrophysical sources is, in practice, biased by the incorrect initial pedestal removal and further iterations should lead to progressively better de-biased data. However, we have found that one iteration is adequate to remove most of the residual bias. This procedure yields four exposures for each SLACS target which have been dark-, amplifier glow-, pedestal-, and sky background-subtracted, and these pointings are then drizzled on to a common output image with a scale of 0\farcs05 using a pixfrac of 0.8.

We also take care in assuring that we have a robust estimate of the variance in this final drizzled image. We convert the \textsf{ERR} frames created by the CALINCA pipeline into variance images by squaring them; these variance images contain an estimate of the noise properties from the dark current and amplifier read noise. The single-exposure variance images are then added to the (non-background-subtracted, pre-drizzle) science images, which have been appropriately scaled by the exposure time per pixel (encoded in the \textsf{TIME} extension created by CALNICA) and the detector gain $g$, to create pixel-to-pixel variance maps for each exposure.
These variance maps are individually drizzled to the output science coordinate frame, yielding single-exposure drizzled variance maps (\textsf{VAR}) and the drizzle `weight' (\textsf{WHT}) images. The \textsf{WHT} images are then used to determine the final variance image associated with the drizzled science image by taking a weighted sum of the \textsf{VAR} images.
We expect this procedure yields the optimal representation of the variance in each pixel of the output (stacked and drizzled) science image.

\section{The Definitive SLACS Sample in Color}
\begin{figure}[ht]
 \centering
 \includegraphics[width=0.48\textwidth,clip]{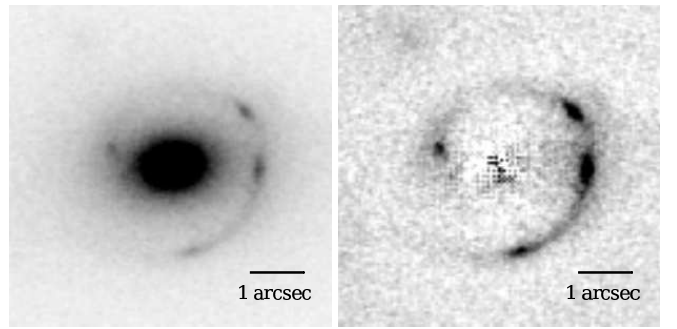}
 \caption{The WFPC2 F606W image of the SLACS candidate SDSSJ2343-0030. The left panel shows the early-type morphology of the lens and evidence for the lensed nature of the background source. The right panel shows the system with B-spline model of the lensing galaxy subtracted, revealing a clear four-image lens system with a partial Einstein ring. See the electronic version of the journal for images of all 12 of the new targets.}
 \label{F_new_targets}
\end{figure}

The SLACS survey has observed 12 new targets (Table \ref{T_new_targets}) since the conclusion of the ACS program (Paper V). The WFPC2 F606W images of these targets are shown in Figure \ref{F_new_targets}, including the residuals from B-spline galaxy model subtractions (see Section 3.1). In addition to these 12 candidates, we revisit 14 systems from Paper V that had grades poorer than `A' for which we have obtained new multi-band imaging. Grade `A' refers to confirmed lenses, while grade `B' denotes probable lenses and grade `X' is used for unlikely lenses; see Paper V for a more detailed description of the grading scheme.

\begin{figure*}[ht]
 \includegraphics[width=0.98\textwidth,clip]{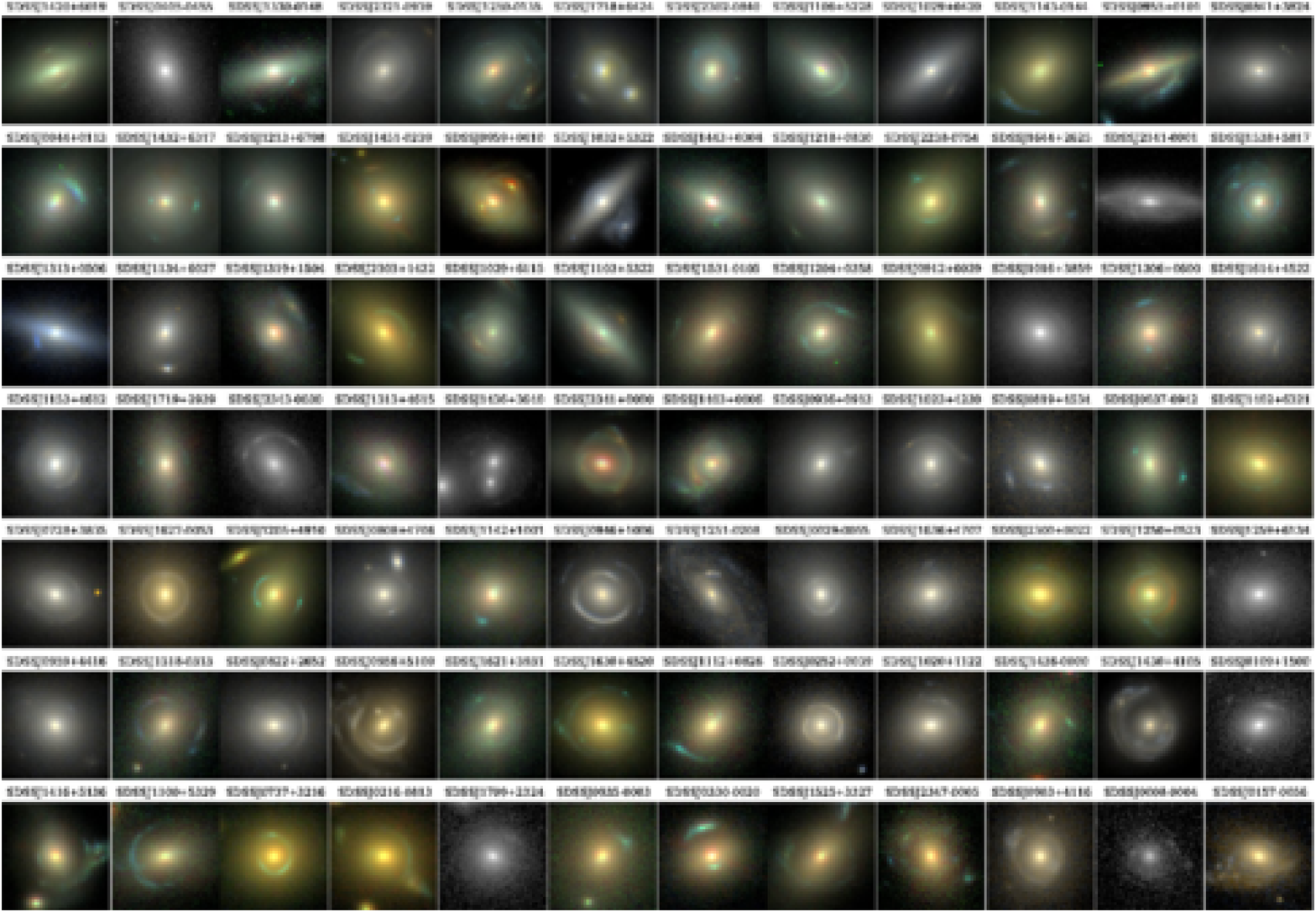}
 \caption{Color cutouts of 84 of the SLACS lenses, ordered by redshift. The lens SDSSJ1618+4353 has two primary lensing galaxies and has not been included. See the Journal version for full-resolution figure.}
 \label{F_color_cutouts}
\end{figure*}

The new NICMOS and WFCP2 data for these systems were treated in a similar manner to the ACS data of Paper V in order to establish the lensing nature of each system. The putative lensing galaxy was subtracted from each image using B-spline models, and the residuals were visually inspected for evidence of lensed images and counter-images. A consensus grade was compiled by four authors (MWA, ASB, LAM, and TT) using the same grading scheme as Paper V. Four of the systems were initially given a `B' grade by the consensus analysis but were subsequently upgraded to `A' when successful lens models were fit to the images (see below). Furthermore, SDSSJ1313+0506 has been confirmed as a lens based upon the identification of a counter-image to the background [\ion{O}{2}] emission from a DEIMOS spectrum (A. Dutton, private communication). In total, 9 of the 12 previously unobserved candidates have been confirmed as `A' lenses, and 6 systems reported on in Paper V have been upgraded to `A' status. This brings the number of grade `A' lens systems identified by SLACS to 85 (Table \ref{T_grade_A_raw}), with an additional 13 `B' candidates. A color montage of the grade `A' lenses can be found in Figure \ref{F_color_cutouts}.

\subsection{Lensing Analysis}
We produce singular isothermal ellipsoid (SIE) lens models \citep[e.g.,][]{kormann} for 11 of the 15 new grade A lenses. One candidate from Paper V, SDSSJ1029+6115, is shown clearly to be a lens using IFU data but our HST data do not have sufficient sensitivity to adequately model the lensing, nor do we model the spectroscopically confirmed lens SDSSJ1313+0506. Additionally, two new candidates from our WFPC2 imaging, SDSSJ1319+1504 and SDSSJ1436+3640, will require special subtractions due to the presence of nearby galaxies in the field.

\begin{figure*}[ht]
 \includegraphics[width=0.98\textwidth,clip]{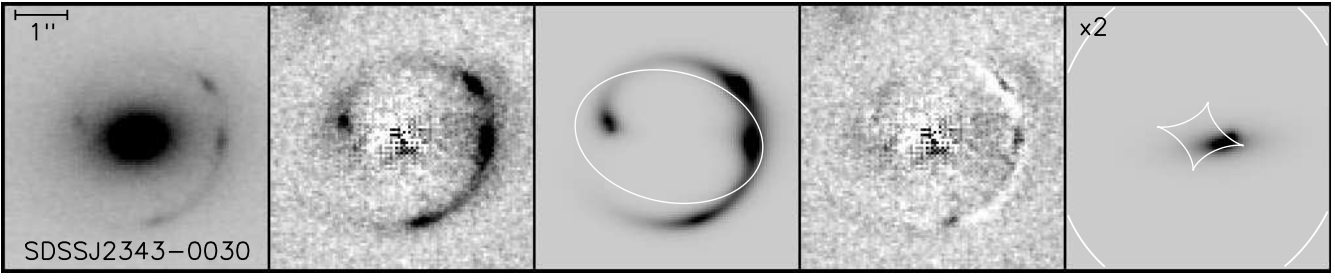}
 \caption{The SIE lens model for the SLACS lens SDSSJ2343-0030. The five panels show: the HST image of the lens; the lensing galaxy-subtracted residuals; the model of the lensed source and the lensing critical curves (in white); the source- and lens-subtracted residuals; and the reconstructed background source and caustics (in white). See the electronic version of the journal for all 11 lens models.}
 \label{F_lens_models}
\end{figure*}

The SIE modeling is performed in the same manner as Paper V, with the only difference being the switch from ACS data to WFPC2 data. To briefly summarize, the mass of the lensing galaxy for each system is modeled as an SIE profile positioned on the center of the light distribution. A lens galaxy-subtracted image of the background source is created by modeling the light of the lens galaxy using a radial B-spline model and subtracting this from our HST images. The background source is modeled as single or multiple Gaussian or Sersic ellipsoid components which are lensed by ray-tracing through the SIE mass model of the lens; this lensed source is fit to the observed images in the lens galaxy-subtracted frame. The model parameters (the Einstein radius and ellipticity parameters for the lens) are first adjusted by hand before optimization by MPFIT \citep{mpfit}. The resulting models are shown in Figure \ref{F_lens_models} and the parameters of the lens models are listed in Table \ref{T_grade_A_model}. We note that these are not perfect lens models (not all of the lens are isothermal, nor are the sources always ideally described by Sersic profiles). However, the SIE models do provide accurate determinations of the Einstein radii and the masses within the Einstein radii; these are the parameters of interest for this paper.

\subsection{The SLACS Catalogs}
The SLACS data products can be split into two categories\footnote{These data will all be available online at http://www.slacs.org}: `basic' data products which include photometric and spectroscopic measurements, and inferred data products, including parameters inferred from lens modeling and stellar population analysis. In Table \ref{T_grade_A_raw} we present our `basic' data for the grade `A' lenses, including the location on the sky, the lens and source redshifts, the lensing galaxy velocity dispersion, the morphological type, and the effective radii (measured at the intermediate axis) and magnitudes from our photometric analysis (Section 4). Table \ref{T_grade_A_model} presents the inferred quantities, including Einstein radii and masses, stellar mass estimates for Chabrier and Salpeter initial mass functions (IMFs) denoted with the subscripts `Chab' and `Salp' respectively, luminosities, and $f_{\rm *,Ein}$, the stellar fraction within the Einstein radius for each IMF.

\LongTables
\begin{deluxetable*}{lllcccrccccccc}
\tabletypesize{\scriptsize}
 \tablecolumns{14}
 \tablewidth{0pc}
 \tablecaption{Photometric and Spectroscopic Properties of SLACS Lenses}
 \tablehead{
   \colhead{Name} &
   \colhead{RA} &
   \colhead{Dec} &
   \colhead{$z_{lens}$} &
   \colhead{$z_{src}$} &
   \colhead{$\sigma_{ap}$} &
   \colhead{Morph} &
   \colhead{m$_B$} &
   \colhead{r$_{e,B}$} &
   \colhead{m$_V$} &
   \colhead{r$_{e,V}$} &
   \colhead{m$_I$} &
   \colhead{r$_{e,I}$} &
   \colhead{m$_H$} \\
   \colhead{}  &
   \colhead{}  &
   \colhead{}  &
   \colhead{}  &
   \colhead{}  &
   \colhead{(km s$^{-1}$)}  &
   \colhead{}  &
   \colhead{}  &
   \colhead{(\arcsec)}  &
   \colhead{}  &
   \colhead{(\arcsec)}  &
   \colhead{}  &
   \colhead{(\arcsec)}  &
   \colhead{}
 }
\startdata
SDSSJ0008$-$0004  &  00h08m02.96s & -00$^{\circ}$04m08.2s  & 0.440  &  1.192  &  193$\pm$36  &        E  &  \nodata &  \nodata &    19.60\tablenotemark{$\dagger$}  &     1.82  &    18.52  &     1.71  &  \nodata \\
SDSSJ0029$-$0055  &  00h29m07.78s & -00$^{\circ}$55m50.5s  & 0.227  &  0.931  &  229$\pm$18  &        E  &  \nodata &  \nodata &    17.74\tablenotemark{$\dagger$}  &     2.58  &    17.05  &     2.16  &  \nodata \\
SDSSJ0037$-$0942  &  00h37m53.21s & -09$^{\circ}$42m20.1s  & 0.195  &  0.632  &  279$\pm$10  &        E  &    18.57 &     2.87 &    16.90\tablenotemark{$\dagger$}  &     2.68  &    16.20  &     1.80  &    15.49 \\
SDSSJ0044$+$0113  &  00h44m02.90s & ~01$^{\circ}$13m12.6s  & 0.120  &  0.197  &  266$\pm$13  &        E  &  \nodata &  \nodata &    16.32\tablenotemark{$\dagger$}  &     3.25  &    15.69  &     1.92  &    15.03 \\
SDSSJ0109$+$1500  &  01h09m33.73s & ~15$^{\circ}$00m32.5s  & 0.294  &  0.525  &  251$\pm$19  &        E  &  \nodata &  \nodata &  \nodata  &  \nodata  &    17.63  &     1.38  &  \nodata \\
SDSSJ0157$-$0056  &  01h57m58.94s & -00$^{\circ}$56m26.1s  & 0.513  &  0.924  &  295$\pm$47  &        E  &  \nodata &  \nodata &    19.87\tablenotemark{$\dagger$}  &     1.21  &    18.59  &     1.84  &    17.52 \\
SDSSJ0216$-$0813  &  02h16m52.54s & -08$^{\circ}$13m45.3s  & 0.332  &  0.523  &  333$\pm$23  &        E  &    19.91 &     2.49 &    18.36\tablenotemark{$\ddagger$}  &     2.97  &    16.86  &     2.40  &    15.99 \\
SDSSJ0252$+$0039  &  02h52m45.21s & ~00$^{\circ}$39m58.4s  & 0.280  &  0.982  &  164$\pm$12  &        E  &  \nodata &  \nodata &    18.77\tablenotemark{$\dagger$}  &     1.36  &    17.87  &     1.39  &  \nodata \\
SDSSJ0330$-$0020  &  03h30m12.14s & -00$^{\circ}$20m51.9s  & 0.351  &  1.071  &  212$\pm$21  &        E  &  \nodata &  \nodata &    18.84\tablenotemark{$\dagger$}  &     1.52  &    17.99  &     0.91  &    17.08 \\
SDSSJ0405$-$0455  &  04h05m35.41s & -04$^{\circ}$55m52.4s  & 0.075  &  0.810  &  160$\pm$ 7  &        E  &  \nodata &  \nodata &  \nodata  &  \nodata  &    16.22  &     1.36  &  \nodata \\
SDSSJ0728$+$3835  &  07h28m04.95s & ~38$^{\circ}$35m25.7s  & 0.206  &  0.688  &  214$\pm$11  &        E  &  \nodata &  \nodata &    17.38\tablenotemark{$\dagger$}  &     2.01  &    16.62  &     1.78  &  \nodata \\
SDSSJ0737$+$3216  &  07h37m28.45s & ~32$^{\circ}$16m18.6s  & 0.322  &  0.581  &  338$\pm$16  &        E  &    20.17 &     2.22 &    18.34\tablenotemark{$\ddagger$}  &     3.38  &    16.95  &     1.80  &    16.17 \\
SDSSJ0808$+$4706  &  08h08m58.78s & ~47$^{\circ}$06m38.9s  & 0.219  &  1.025  &  236$\pm$11  &        E  &  \nodata &  \nodata &    17.11\tablenotemark{$\dagger$}  &     4.53  &    16.74  &     2.42  &  \nodata \\
SDSSJ0819$+$4534  &  08h19m31.93s & ~45$^{\circ}$34m44.8s  & 0.194  &  0.446  &  225$\pm$15  &        E  &    19.16 &     2.91 &    17.61\tablenotemark{$\dagger$}  &     2.80  &    16.99  &     1.98  &  \nodata \\
SDSSJ0822$+$2652  &  08h22m42.32s & ~26$^{\circ}$52m43.5s  & 0.241  &  0.594  &  259$\pm$15  &        E  &  \nodata &  \nodata &    17.65\tablenotemark{$\dagger$}  &     2.43  &    16.94  &     1.82  &  \nodata \\
SDSSJ0841$+$3824  &  08h41m28.81s & ~38$^{\circ}$24m13.7s  & 0.116  &  0.657  &  225$\pm$ 8  &        S  &  \nodata &  \nodata &    15.66\tablenotemark{$\dagger$}  &     8.67  &    15.27  &     4.21  &  \nodata \\
SDSSJ0903$+$4116  &  09h03m15.19s & ~41$^{\circ}$16m09.1s  & 0.430  &  1.065  &  223$\pm$27  &        E  &  \nodata &  \nodata &    18.90\tablenotemark{$\dagger$}  &     2.19  &    17.91  &     1.78  &  \nodata \\
SDSSJ0912$+$0029  &  09h12m05.31s & ~00$^{\circ}$29m01.2s  & 0.164  &  0.324  &  326$\pm$12  &     E/S0  &    17.91 &     4.51 &    16.56\tablenotemark{$\ddagger$}  &     4.29  &    15.52  &     4.01  &    14.52 \\
SDSSJ0935$-$0003  &  09h35m43.93s & -00$^{\circ}$03m34.8s  & 0.347  &  0.467  &  396$\pm$35  &        E  &    19.36 &     5.59 &    17.71\tablenotemark{$\dagger$}  &     4.12  &    16.69  &     2.15  &    16.15 \\
SDSSJ0936$+$0913  &  09h36m00.77s & ~09$^{\circ}$13m35.8s  & 0.190  &  0.588  &  243$\pm$11  &        E  &  \nodata &  \nodata &    17.12\tablenotemark{$\dagger$}  &     2.50  &    16.44  &     2.11  &  \nodata \\
SDSSJ0946$+$1006  &  09h46m56.68s & ~10$^{\circ}$06m52.8s  & 0.222  &  0.609  &  263$\pm$21  &        E  &  \nodata &  \nodata &    17.78\tablenotemark{$\dagger$}  &     2.77  &    17.04  &     2.35  &  \nodata \\
SDSSJ0955$+$0101  &  09h55m19.72s & ~01$^{\circ}$01m44.4s  & 0.111  &  0.316  &  192$\pm$13  &        S  &    19.16 &     1.82 &    17.65\tablenotemark{$\dagger$}  &     1.98  &    16.92  &     1.47  &    16.03 \\
SDSSJ0956$+$5100  &  09h56m29.78s & ~51$^{\circ}$00m06.6s  & 0.241  &  0.470  &  334$\pm$15  &        E  &    19.26 &     2.54 &    17.90\tablenotemark{$\ddagger$}  &     2.32  &    16.66  &     2.19  &  \nodata \\
SDSSJ0959$+$4416  &  09h59m00.96s & ~44$^{\circ}$16m39.4s  & 0.237  &  0.531  &  244$\pm$19  &       S0  &  \nodata &  \nodata &    17.67\tablenotemark{$\dagger$}  &     2.04  &    16.88  &     1.98  &  \nodata \\
SDSSJ0959$+$0410  &  09h59m44.07s & ~04$^{\circ}$10m17.0s  & 0.126  &  0.535  &  197$\pm$13  &        E  &    19.14 &     1.56 &    17.94\tablenotemark{$\ddagger$}  &     1.51  &    16.87  &     1.29  &    16.01 \\
SDSSJ1016$+$3859  &  10h16m22.86s & ~38$^{\circ}$59m03.3s  & 0.168  &  0.439  &  247$\pm$13  &        E  &  \nodata &  \nodata &    17.35\tablenotemark{$\dagger$}  &     1.65  &    16.67  &     1.46  &  \nodata \\
SDSSJ1020$+$1122  &  10h20m26.54s & ~11$^{\circ}$22m41.1s  & 0.282  &  0.553  &  282$\pm$18  &        E  &  \nodata &  \nodata &    18.12\tablenotemark{$\dagger$}  &     1.52  &    17.15  &     1.59  &  \nodata \\
SDSSJ1023$+$4230  &  10h23m32.26s & ~42$^{\circ}$30m01.8s  & 0.191  &  0.696  &  242$\pm$15  &        E  &  \nodata &  \nodata &    17.45\tablenotemark{$\dagger$}  &     2.07  &    16.74  &     1.77  &  \nodata \\
SDSSJ1029$+$0420  &  10h29m22.94s & ~04$^{\circ}$20m01.8s  & 0.104  &  0.615  &  210$\pm$ 9  &       S0  &  \nodata &  \nodata &    16.69\tablenotemark{$\dagger$}  &     2.15  &    16.06  &     1.56  &  \nodata \\
SDSSJ1029$+$6115  &  10h29m27.53s & ~61$^{\circ}$15m05.3s  & 0.157  &  0.251  &  228$\pm$14  &        E  &  \nodata &  \nodata &    16.82\tablenotemark{$\dagger$}  &     2.75  &    16.04  &     2.52  &    15.04 \\
SDSSJ1032$+$5322  &  10h32m35.84s & ~53$^{\circ}$22m34.9s  & 0.133  &  0.329  &  296$\pm$14  &        S  &  \nodata &  \nodata &    17.76\tablenotemark{$\dagger$}  &     1.19  &    17.01  &     0.81  &  \nodata \\
SDSSJ1100$+$5329  &  11h00m24.39s & ~53$^{\circ}$29m13.9s  & 0.317  &  0.858  &  187$\pm$23  &        E  &    19.78 &     2.90 &    18.06\tablenotemark{$\dagger$}  &     2.96  &    17.15  &     2.20  &    16.26 \\
SDSSJ1103$+$5322  &  11h03m08.21s & ~53$^{\circ}$22m28.2s  & 0.158  &  0.735  &  196$\pm$12  &    S0/SA  &  \nodata &  \nodata &    17.14\tablenotemark{$\dagger$}  &     2.61  &    16.41  &     2.85  &    15.54 \\
SDSSJ1106$+$5228  &  11h06m46.15s & ~52$^{\circ}$28m37.8s  & 0.095  &  0.407  &  262$\pm$ 9  &        E  &    17.54 &     1.79 &    16.08\tablenotemark{$\dagger$}  &     2.54  &    15.50  &     1.39  &    14.81 \\
SDSSJ1112$+$0826  &  11h12m50.60s & ~08$^{\circ}$26m10.4s  & 0.273  &  0.629  &  320$\pm$20  &        E  &  \nodata &  \nodata &    17.97\tablenotemark{$\dagger$}  &     1.83  &    17.16  &     1.32  &    16.21 \\
SDSSJ1134$+$6027  &  11h34m05.89s & ~60$^{\circ}$27m13.5s  & 0.153  &  0.474  &  239$\pm$11  &        E  &  \nodata &  \nodata &    17.14\tablenotemark{$\dagger$}  &     2.15  &    16.41  &     2.02  &  \nodata \\
SDSSJ1142$+$1001  &  11h42m57.35s & ~10$^{\circ}$01m11.8s  & 0.222  &  0.504  &  221$\pm$22  &        E  &  \nodata &  \nodata &    17.75\tablenotemark{$\dagger$}  &     2.11  &    16.99  &     1.24  &    16.23 \\
SDSSJ1143$-$0144  &  11h43m29.64s & -01$^{\circ}$44m30.0s  & 0.106  &  0.402  &  269$\pm$ 5  &        E  &  \nodata &  \nodata &    15.83\tablenotemark{$\ddagger$}  &     5.41  &    14.92  &     2.66  &    14.40 \\
SDSSJ1153$+$4612  &  11h53m10.79s & ~46$^{\circ}$12m05.3s  & 0.180  &  0.875  &  226$\pm$15  &        E  &  \nodata &  \nodata &    17.78\tablenotemark{$\dagger$}  &     1.43  &    17.15  &     1.16  &  \nodata \\
SDSSJ1204$+$0358  &  12h04m44.07s & ~03$^{\circ}$58m06.4s  & 0.164  &  0.631  &  267$\pm$17  &        E  &    19.33 &     1.43 &    17.45\tablenotemark{$\dagger$}  &     1.65  &    16.80  &     1.09  &    15.89 \\
SDSSJ1205$+$4910  &  12h05m40.44s & ~49$^{\circ}$10m29.4s  & 0.215  &  0.481  &  281$\pm$13  &        E  &    19.19 &     2.61 &    17.70\tablenotemark{$\ddagger$}  &     2.59  &    16.52  &     1.79  &    15.82 \\
SDSSJ1213$+$6708  &  12h13m40.58s & ~67$^{\circ}$08m29.0s  & 0.123  &  0.640  &  292$\pm$11  &        E  &  \nodata &  \nodata &    16.28\tablenotemark{$\dagger$}  &     3.36  &    15.57  &     1.50  &    15.05 \\
SDSSJ1218$+$0830  &  12h18m26.70s & ~08$^{\circ}$30m50.3s  & 0.135  &  0.717  &  219$\pm$10  &        E  &  \nodata &  \nodata &    16.38\tablenotemark{$\dagger$}  &     3.77  &    15.71  &     2.70  &    15.02 \\
SDSSJ1250$+$0523  &  12h50m28.26s & ~05$^{\circ}$23m49.1s  & 0.232  &  0.795  &  252$\pm$14  &        E  &    19.19 &     1.93 &    17.77\tablenotemark{$\ddagger$}  &     1.91  &    16.64  &     1.32  &    15.78 \\
SDSSJ1250$-$0135  &  12h50m50.52s & -01$^{\circ}$35m31.7s  & 0.087  &  0.353  &  246$\pm$ 9  &    S0/SA  &    16.89 &     4.56 &    15.68\tablenotemark{$\dagger$}  &     4.01  &    15.09  &     2.03  &    14.48 \\
SDSSJ1251$-$0208  &  12h51m35.70s & -02$^{\circ}$08m05.2s  & 0.224  &  0.784  &  233$\pm$23  &        S  &    19.54 &     3.61 &    17.71\tablenotemark{$\dagger$}  &     5.34  &    17.21  &     2.61  &  \nodata \\
SDSSJ1259$+$6134  &  12h59m19.05s & ~61$^{\circ}$34m08.6s  & 0.233  &  0.449  &  253$\pm$16  &        E  &    19.56 &     1.78 &  \nodata  &  \nodata  &    16.83  &     1.81  &  \nodata \\
SDSSJ1306$+$0600  &  13h06m13.65s & ~06$^{\circ}$00m22.1s  & 0.173  &  0.472  &  237$\pm$17  &        E  &  \nodata &  \nodata &    17.39\tablenotemark{$\dagger$}  &     2.34  &    16.76  &     1.25  &    15.96 \\
SDSSJ1313$+$4615  &  13h13m02.93s & ~46$^{\circ}$15m13.6s  & 0.185  &  0.514  &  263$\pm$18  &        E  &  \nodata &  \nodata &    17.29\tablenotemark{$\dagger$}  &     2.15  &    16.63  &     1.59  &    15.73 \\
SDSSJ1313$+$0506  &  13h13m26.70s & ~05$^{\circ}$06m57.2s  & 0.144  &  0.339  &  221$\pm$17  &        S  &  \nodata &  \nodata &    17.71\tablenotemark{$\dagger$}  &     1.53  &    17.03  &     0.97  &    16.25 \\
SDSSJ1318$-$0313  &  13h18m39.33s & -03$^{\circ}$13m34.2s  & 0.240  &  1.300  &  213$\pm$18  &        E  &  \nodata &  \nodata &    17.54\tablenotemark{$\dagger$}  &     4.18  &    16.80  &     2.51  &    16.06 \\
SDSSJ1319$+$1504  &  13h19m32.50s & ~15$^{\circ}$04m26.0s  & 0.154  &  0.606  &  235$\pm$14  &     E/S0  &  \nodata &  \nodata &    17.89\tablenotemark{$\dagger$}  &     1.44  &    17.23  &     0.77  &    16.32 \\
SDSSJ1330$-$0148  &  13h30m45.53s & -01$^{\circ}$48m41.6s  & 0.081  &  0.711  &  185$\pm$ 9  &       S0  &    18.91 &     1.00 &    17.56\tablenotemark{$\dagger$}  &     1.36  &    16.92  &     0.96  &    16.18 \\
SDSSJ1402$+$6321  &  14h02m28.21s & ~63$^{\circ}$21m33.5s  & 0.205  &  0.481  &  267$\pm$17  &        E  &    19.07 &     2.56 &    17.51\tablenotemark{$\ddagger$}  &     2.62  &    16.30  &     2.29  &    15.55 \\
SDSSJ1403$+$0006  &  14h03m29.49s & ~00$^{\circ}$06m41.4s  & 0.189  &  0.473  &  213$\pm$17  &        E  &  \nodata &  \nodata &    17.64\tablenotemark{$\dagger$}  &     1.87  &    17.02  &     1.14  &    16.11 \\
SDSSJ1416$+$5136  &  14h16m22.34s & ~51$^{\circ}$36m30.4s  & 0.299  &  0.811  &  240$\pm$25  &        E  &  \nodata &  \nodata &    18.58\tablenotemark{$\dagger$}  &     1.38  &    17.55  &     0.98  &    16.70 \\
SDSSJ1420$+$6019  &  14h20m15.85s & ~60$^{\circ}$19m14.8s  & 0.063  &  0.535  &  205$\pm$ 4  &       S0  &    16.88 &     2.35 &    15.94\tablenotemark{$\ddagger$}  &     2.12  &    15.05  &     2.25  &    14.37 \\
SDSSJ1430$+$4105  &  14h30m04.10s & ~41$^{\circ}$05m57.1s  & 0.285  &  0.575  &  322$\pm$32  &        E  &  \nodata &  \nodata &    17.80\tablenotemark{$\dagger$}  &     2.71  &    16.85  &     2.55  &  \nodata \\
SDSSJ1432$+$6317  &  14h32m13.34s & ~63$^{\circ}$17m03.8s  & 0.123  &  0.664  &  199$\pm$ 8  &        S  &  \nodata &  \nodata &    15.78\tablenotemark{$\dagger$}  &     6.21  &    15.13  &     3.04  &    14.53 \\
SDSSJ1436$+$3640  &  14h36m05.63s & ~36$^{\circ}$40m47.1s  & 0.185  &  0.758  &  281$\pm$19  &  E\tablenotemark{$ast$}  &  \nodata &  \nodata &    16.95\tablenotemark{$\dagger$}  &     4.64  &  \nodata  &  \nodata  &  \nodata \\
SDSSJ1436$-$0000  &  14h36m27.54s & -00$^{\circ}$00m29.2s  & 0.285  &  0.805  &  224$\pm$17  &        E  &  \nodata &  \nodata &    17.94\tablenotemark{$\dagger$}  &     2.94  &    17.16  &     1.63  &    16.38 \\
SDSSJ1443$+$0304  &  14h43m19.62s & ~03$^{\circ}$04m08.2s  & 0.134  &  0.419  &  209$\pm$11  &       S0  &    19.21 &     1.13 &    17.62\tablenotemark{$\dagger$}  &     1.38  &    17.00  &     0.70  &    16.24 \\
SDSSJ1451$-$0239  &  14h51m28.19s & -02$^{\circ}$39m36.4s  & 0.125  &  0.520  &  223$\pm$14  &        E  &    17.88 &     3.49 &    16.92\tablenotemark{$\ddagger$}  &     2.64  &    15.92  &     1.54  &    15.27 \\
SDSSJ1525$+$3327  &  15h25m06.70s & ~33$^{\circ}$27m47.4s  & 0.358  &  0.717  &  264$\pm$26  &        E  &  \nodata &  \nodata &    18.10\tablenotemark{$\dagger$}  &     3.40  &    17.07  &     2.42  &    16.09 \\
SDSSJ1531$-$0105  &  15h31m50.07s & -01$^{\circ}$05m45.7s  & 0.160  &  0.744  &  279$\pm$12  &        E  &  \nodata &  \nodata &    16.43\tablenotemark{$\dagger$}  &     3.32  &    15.79  &     1.97  &    15.14 \\
SDSSJ1538$+$5817  &  15h38m12.92s & ~58$^{\circ}$17m09.8s  & 0.143  &  0.531  &  189$\pm$12  &        E  &  \nodata &  \nodata &    17.42\tablenotemark{$\dagger$}  &     1.59  &    16.63  &     1.00  &    15.99 \\
SDSSJ1614$+$4522  &  16h14m37.74s & ~45$^{\circ}$22m53.3s  & 0.178  &  0.811  &  182$\pm$13  &        E  &  \nodata &  \nodata &    17.50\tablenotemark{$\dagger$}  &     2.93  &    16.80  &     2.58  &  \nodata \\
SDSSJ1618$+$4353  &  16h18m43.10s & ~43$^{\circ}$53m27.4s  & 0.199  &  0.666  &  292$\pm$29  &  E\tablenotemark{$ast$}  &    19.27 &     2.53 &  \nodata  &  \nodata  &    17.08  &     1.80  &  \nodata \\
SDSSJ1621$+$3931  &  16h21m32.99s & ~39$^{\circ}$31m44.6s  & 0.245  &  0.602  &  236$\pm$20  &        E  &    19.27 &     2.85 &    17.55\tablenotemark{$\dagger$}  &     2.77  &    16.79  &     1.51  &    16.08 \\
SDSSJ1627$-$0053  &  16h27m46.45s & -00$^{\circ}$53m57.6s  & 0.208  &  0.524  &  290$\pm$14  &        E  &    19.42 &     2.02 &    17.87\tablenotemark{$\ddagger$}  &     2.05  &    16.71  &     1.98  &  \nodata \\
SDSSJ1630$+$4520  &  16h30m28.16s & ~45$^{\circ}$20m36.3s  & 0.248  &  0.793  &  276$\pm$16  &        E  &    19.76 &     1.81 &    18.08\tablenotemark{$\ddagger$}  &     2.05  &    16.77  &     1.65  &    15.80 \\
SDSSJ1636$+$4707  &  16h36m02.62s & ~47$^{\circ}$07m29.6s  & 0.228  &  0.675  &  231$\pm$15  &        E  &    19.62 &     1.76 &    17.79\tablenotemark{$\dagger$}  &     1.84  &    16.98  &     1.68  &  \nodata \\
SDSSJ1644$+$2625  &  16h44m43.09s & ~26$^{\circ}$25m25.4s  & 0.137  &  0.610  &  229$\pm$12  &        E  &  \nodata &  \nodata &    16.85\tablenotemark{$\dagger$}  &     2.32  &    16.22  &     1.55  &    15.46 \\
SDSSJ1709$+$2324  &  17h09m38.96s & ~23$^{\circ}$24m08.4s  & 0.347  &  0.719  &  286$\pm$30  &        E  &  \nodata &  \nodata &    18.51\tablenotemark{$\dagger$}  &     2.06  &  \nodata  &  \nodata  &  \nodata \\
SDSSJ1718$+$6424  &  17h18m37.39s & ~64$^{\circ}$24m52.2s  & 0.090  &  0.737  &  273$\pm$16  &  E\tablenotemark{$ast$}  &    17.39 &     4.69 &    16.15\tablenotemark{$\dagger$}  &     4.51  &    15.39  &     1.70  &    15.35 \\
SDSSJ1719$+$2939  &  17h19m34.16s & ~29$^{\circ}$39m26.4s  & 0.181  &  0.578  &  286$\pm$15  &     E/S0  &  \nodata &  \nodata &    17.52\tablenotemark{$\dagger$}  &     1.83  &    16.87  &     1.46  &    15.99 \\
SDSSJ2141$-$0001  &  21h41m54.68s & -00$^{\circ}$01m12.3s  & 0.138  &  0.713  &  181$\pm$14  &        S  &  \nodata &  \nodata &  \nodata  &  \nodata  &    16.72  &     1.81  &  \nodata \\
SDSSJ2238$-$0754  &  22h38m40.20s & -07$^{\circ}$54m56.0s  & 0.137  &  0.713  &  198$\pm$11  &        E  &    18.21 &     3.19 &    17.18\tablenotemark{$\ddagger$}  &     2.41  &    16.12  &     1.82  &    15.40 \\
SDSSJ2300$+$0022  &  23h00m53.15s & ~00$^{\circ}$22m38.0s  & 0.228  &  0.463  &  279$\pm$17  &        E  &    19.80 &     1.86 &    18.19\tablenotemark{$\ddagger$}  &     1.93  &    16.96  &     1.52  &    16.15 \\
SDSSJ2302$-$0840  &  23h02m20.18s & -08$^{\circ}$40m49.5s  & 0.090  &  0.222  &  237$\pm$ 8  &        E  &  \nodata &  \nodata &    15.95\tablenotemark{$\dagger$}  &     3.18  &    15.45  &     1.70  &    14.80 \\
SDSSJ2303$+$1422  &  23h03m21.72s & ~14$^{\circ}$22m17.9s  & 0.155  &  0.517  &  255$\pm$16  &        E  &    17.96 &     4.28 &    16.77\tablenotemark{$\ddagger$}  &     3.54  &    15.72  &     2.94  &    15.05 \\
SDSSJ2321$-$0939  &  23h21m20.93s & -09$^{\circ}$39m10.3s  & 0.082  &  0.532  &  249$\pm$ 8  &        E  &    16.52 &     5.54 &    15.27\tablenotemark{$\dagger$}  &     4.79  &    14.61  &     4.11  &  \nodata \\
SDSSJ2341$+$0000  &  23h41m11.57s & ~00$^{\circ}$00m18.7s  & 0.186  &  0.807  &  207$\pm$13  &        E  &  \nodata &  \nodata &    17.02\tablenotemark{$\dagger$}  &     4.52  &    16.30  &     2.36  &    15.40 \\
SDSSJ2343$-$0030  &  23h43m58.87s & -00$^{\circ}$30m22.4s  & 0.181  &  0.463  &  269$\pm$11  &     E/S0  &  \nodata &  \nodata &    17.17\tablenotemark{$\dagger$}  &     2.74  &  \nodata  &  \nodata  &  \nodata \\
SDSSJ2347$-$0005  &  23h47m28.08s & -00$^{\circ}$05m21.3s  & 0.417  &  0.714  &  404$\pm$59  &        E  &    20.73 &     1.81 &    18.82\tablenotemark{$\dagger$}  &     1.80  &    17.83  &     1.14  &    16.85 \\
\vspace{-7pt}

\enddata
\label{T_grade_A_raw}
\tablenotetext{$\dagger$}{From the F606W filter on WFPC2.}
\tablenotetext{$\ddagger$}{From the F555W filter on ACS.}
\tablenotetext{$\ast$}{The lens galaxy has at least one close companion galaxy.}
\end{deluxetable*}


\section{Model Photometry}
The program GALFIT \citep{galfit} is used to perform de Vaucouleurs model fitting to the WFPC2 and NICMOS images of the lenses; we adopt the model photometry using the method from Paper V for the ACS imaging. Subimages of 30\arcsec\ on a side are used to fit the WFPC2 data, and we use the region of the NICMOS mosaic that was covered in all four pointings (generally an area approximately $15\arcsec\times15\arcsec$). Masks of neighboring galaxies and the flux from any lensed images are manually created by iteratively modeling and subtracting the lens galaxy. Bright galaxies very close to the primary lensing galaxy are simultaneously modeled with GALFIT to accurately separate the two objects. We test the robustness of the de Vaucouleurs model magnitudes obtained with GALFIT by comparing the results for lenses observed multiple times and by comparing with magnitudes from a Sersic fit with the Sersic index left as a free parameter; we find that the de Vaucouleurs magnitudes are consistent with a scatter of $\sim0.03$ magnitudes. We also record the effective radii, $r_e$, measured at the intermediate axis for the WFPC2 data and impose a typical model error of 3.5\% (e.g., Paper V; the formal statistical errors are much smaller).

We perform synthetic photometry using composite stellar population (CSP) models (see Section 5); modeling was performed separately for the HST photometry and the SDSS photometry, but we focus on the HST photometry here. The several HST colors available for most systems allow us to constrain the restframe V-band luminosity quite well, and the restframe B-band magnitude with slightly larger uncertainty. The CSP models are also evolved to $z = 0$ to determine self-consistent evolved luminosities in the B and V bands; a more detailed look at the luminosity evolution will be presented in Paper X, but we note that the evolution inferred from the CSP models is in good agreement with other methods (Paper II, G09). Furthermore, we use the HST photometry and CSP models to perform synthetic photometry of the SDSS bands and compare these with the measured SDSS magnitudes for all systems with at least 3 HST bands (56 systems in total). Our comparison is with de-reddened de Vaucouleurs magnitudes (the deVMag magnitudes from the SDSS database corrected for Galactic extinction) in the bands $g,r,i,z$. We find that we are able to reproduce the SDSS photometry quite well, with typical random differences of only a few hundredths of a magnitude. To investigate the affect of the source light on SDSS photometry, we degrade our HST data to the resolution of the SDSS data (approximately 1\farcs5) and perform the photometric modeling on these data (this is done without adding additional noise; we therefore are only investigating the contribution of the background source in low-resolution imaging and without explicit masks during the modeling). We find that we are able to recover the full-resolution magnitudes quite well; in general the sources are too faint to significantly affect the lens light photometry.

\begin{figure}[ht]
 \centering
 \includegraphics[width=0.48\textwidth,clip]{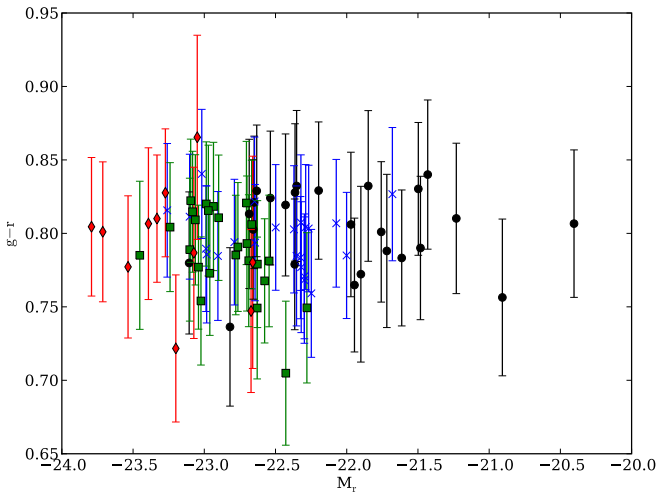}
 \caption{The restframe color-magnitude diagram for SLACS lenses. Black circles denote lenses at $z < 0.15$, blue crosses indicate lenses with redshift $0.15 < z < 0.2$, green squares are lenses at $0.2 < z < 0.25$, and red diamonds are for lenses with $z > 0.25$. Typical early-type galaxies in the local universe have colors ${\rm g-r} \sim0.8$.}
 \label{F_color_magnitude}
\end{figure}

\begin{figure}[ht]
 \centering
 \includegraphics[width=0.48\textwidth,clip]{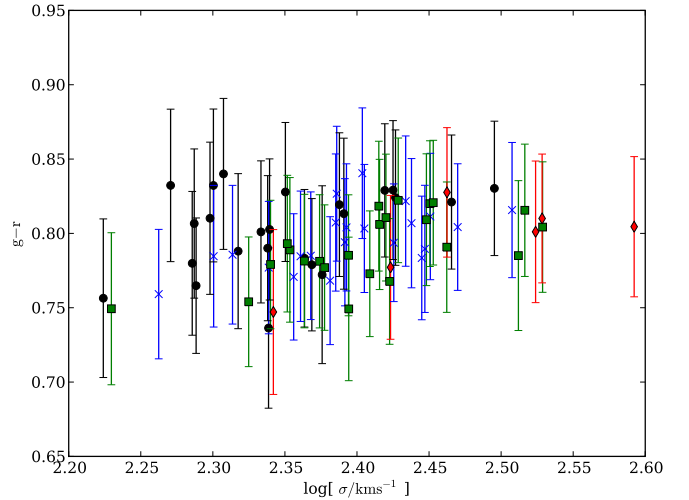}
 \caption{The restframe color-$\sigma$ diagram for SLACS lenses. The color/symbol coding is the same as in Figure \ref{F_color_magnitude}.}
 \label{F_color_sigma}
\end{figure}

We have not determined aperture colors from the HST photometry for the SLACS lenses (all of our HST photometry is from full de Vaucouleurs model fits), but as a sanity check we use the SDSS photometric data to determine the restframe $g-r$ aperture colors \citep[here we use the SDSS modelMag magnitudes to determine the colors, e.g.,][]{stoughton}. As is typical of early-type galaxies, we find a tight sequence with $g-r \sim0.8$, and we plot a restframe color-magnitude diagram for the lenses in Figure \ref{F_color_magnitude}. The absence of any significant tilt in the color-magnitude relation is due to the narrow effective velocity dispersion range of the SLACS lenses \citep[comparable in width to the `constant velocity dispersion' bins used by other authors, e.g.,][]{bernardi,graves}; a slightly clearer trend is seen in a color-$\sigma$ diagram (Figure \ref{F_color_sigma}).

\section{Stellar Masses and Stellar Populations}
One of the key benefits of having multi-color imaging is the ability to determine robust stellar mass estimates, particularly when near-infrared bands are included. The stellar masses of a subset of SLACS lenses have previously been investigated using the ground-based SDSS imaging \citep[G09;][]{grillo}, but the HST-based photometric data are substantially deeper, provide significantly better angular resolution, and extend into the near infrared for most objects.

We have developed a novel code to perform a Bayesian exploration of the stellar populations of galaxies using composite stellar population models produced with third-party codes \citep[e.g.,][hereafter referred to as BC03]{bc03}. The code is described in detail in the Appendix, where we test the robustness of the code and the effects of parameter degeneracies and different filter combinations, but we briefly present the salient features here. The code takes a set of photometric data (magnitudes and errors through given bands) for a galaxy at a known redshift and creates a set of interpolation models for the synthetic magnitudes derived from a grid of CSP model spectra. Because the CSP models can be evaluated at arbitrary points within the bounds of the grid, the user can place arbitrary priors on the CSP model parameters. The parameter space is then explored using a Markov chain Monte Carlo (MCMC) routine, allowing a full determination of the posterior probability distribution function.

In this paper, we restrict ourselves to using BC03 to create our CSP models, employing Chabrier or Salpeter IMFs and an exponential-tau model; the choice of models and IMF will be discussed elsewhere (Treu et al. 2009, in preparation). This star formation history model includes five free parameters: the time that has passed since the beginning of the star formation ($t$), the characteristic time scale of the exponential burst ($\tau$), metallicity ($Z$), reddening (via the $\tau_V$ parameter of BC03), and the stellar mass (M$_*$; see Table 5). We place a uniform prior on $t$ such that the star formation begins between redshifts $z = 1$ and $z = 5$. An exponential prior is used for $\tau$ with scale length 1 Gyr, and we use an extinction prior that is uniform in log$[\tau_V]$. There is a known velocity dispersion-metallicity correlation, and we use the sample of \citet{gallazzi} to create a Gaussian prior on log$[Z]$; this is done by finding the mean and standard deviation on log$[Z]$ for all of the galaxies from \citet{gallazzi} with similar redshift and velocity dispersion to the lensing galaxy.

\begin{deluxetable}{lll}
\tablenum 5
\tabletypesize{\scriptsize}
\tablecolumns{3}
\tablewidth{0pc}
\tablecaption{Stellar Mass Priors}
\tablehead{
 \colhead{Parameter} &
 \colhead{Type} &
 \colhead{Range}
}
\startdata
    M$_*$    &  Log-Uniform  &  $10^9$ to $10^{13}$ M$_\odot$    \\
    $t$      &  Uniform      &  $t_{z = 1}$ to $t_{z = 5}$  \\
    $\tau$   &  Exponential  &  scale length of 1 Gyr      \\
    $\tau_V$ &  Log-Uniform  &  0.01 to 2.0            \\
    $Z$      &  Log-Normal   &  See text.  
\enddata
\label{T_stellar_mass_priors}
\end{deluxetable}

We apply this code to the HST photometry described in Section 4 and Table \ref{T_grade_A_raw}. The results of our analysis are listed in Table \ref{T_grade_A_model}, where we report the mean and standard deviation of the marginalized posterior for the stellar mass for all systems with at least 2 HST bands; we also report the inferred stellar mass fraction within the Einstein radius ($f_{\rm *,Ein}$) for lenses with SIE models. Note that the stellar mass estimates do not include any priors from the lensing, and the inferred mass fractions within the Einstein radius may therefore exceed the total mass inferred from lensing (i.e., $f_{\rm *,Ein}$ may be greater than 1). We additionally provide the ratio of the fraction of light within the Einstein radius and the effective radius to allow easy conversion to the stellar fraction within the effective radius; note that this conversion implies an extrapolation of the lensing mass to larger radii under the assumption of an isothermal mass density profile.

\begin{figure}[ht]
 \centering
 \includegraphics[width=0.48\textwidth,clip]{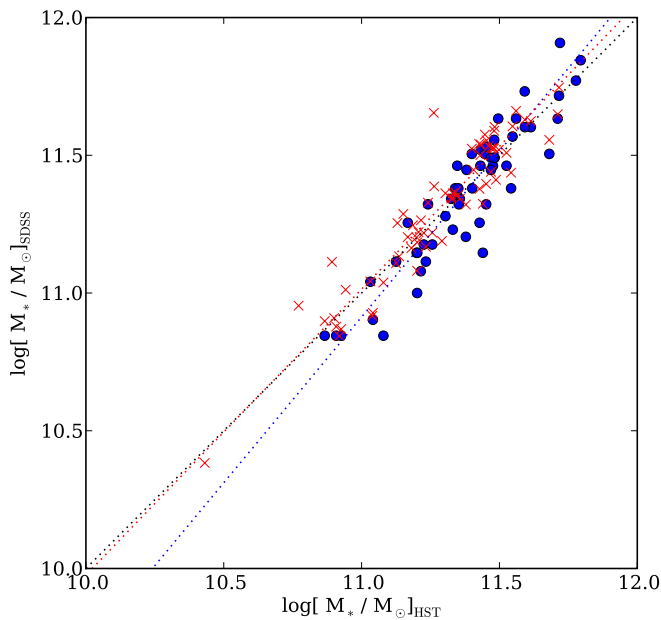}
 \caption{Comparison between stellar masses determined using the SDSS photometry and using the HST photometry. The red crosses are a comparison with the masses determined by the MPA/JHU group using SDSS photometry while the blue circles are a comparison with the stellar masses determined by G09 from SDSS photometry. Note that G09 tend to under-estimate the stellar masses of lower mass lenses compared to the MPA/JHU group and the masses from HST photometry. The red dotted line is a linear fit to the relation between MPA/JHU and HST masses, the blue dotted line is a linear fit to the relation between G09 and HST masses, and the black dotted line indicates the identity.}
 \label{F_mass_photometry_comparisons}
\end{figure}

We have compared the stellar mass estimates derived using the HST photometry with stellar mass estimates from SDSS photometry. We first compare with the masses determined by the MPA/JHU group\footnote{Available at http://www.mpa-garching.mpg.de/SDSS/.}. This sample includes 68 of the systems with more than one band of HST imaging, and we find that the stellar mass estimates from the SDSS data are in very good agreement with our mass estimates using HST photometry (Figure \ref{F_mass_photometry_comparisons}). We also use our stellar mass estimation code to determine masses from the SDSS photometry and find that we are able to reproduce the HST masses and the MPA/JHU masses.

G09 estimate the stellar masses of a subsample of 57 of the SLACS lenses using four different CSP models, including a BC03 delayed exponential-tau model with metallicity set to solar metallicity and without modeling the effects of dust. Figure \ref{F_mass_photometry_comparisons} shows the comparison between the G09 stellar mass estimates and our stellar mass estimates for 55 of the systems investigated by G09 (we only have single band HST imaging for two of the systems in G09). We find a slight mass-dependent bias between the G09 stellar masses and those derived from our HST imaging; the G09 estimates for lower mass lenses are systematically below the masses estimated from HST imaging. We note that galaxies with velocity dispersions similar to the SLACS lenses generally have super-solar metallicities \citep[e.g.,][]{gallazzi} and the G09 choice of solar metallicity may therefore be inappropriate; we find a range of metallicities, including some sub-solar, in our modeling. Differences in priors on the age-determining parameters ($t$ and $\tau$) could also account for the offset seen between the G09 estimates of M$_*$ and the HST and MPA/JHU masses.

\section{The Properties of SLACS Lenses}
\begin{figure}[ht]
 \centering
 \includegraphics[width=0.48\textwidth,clip]{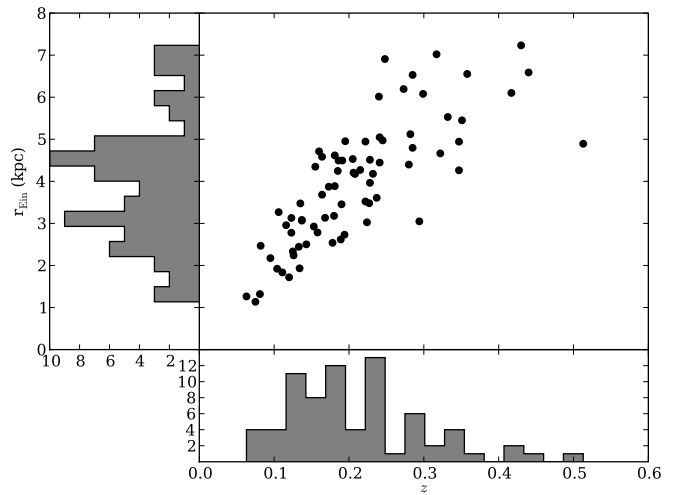}
 \caption{The redshift and Einstein radius, $r_{\rm Ein}$, distributions for SLACS lenses. The median redshift is 0.19 and the median $r_{\rm Ein}$ is 4 kpc. There is a strong correlation between mass and redshift that results from the magnitude limit of the SDSS spectroscopic survey (see Figure \ref{F_redshift_mag}).}
 \label{F_redshift_rein}
\end{figure}

\begin{figure}[ht]
 \centering
 \includegraphics[width=0.48\textwidth,clip]{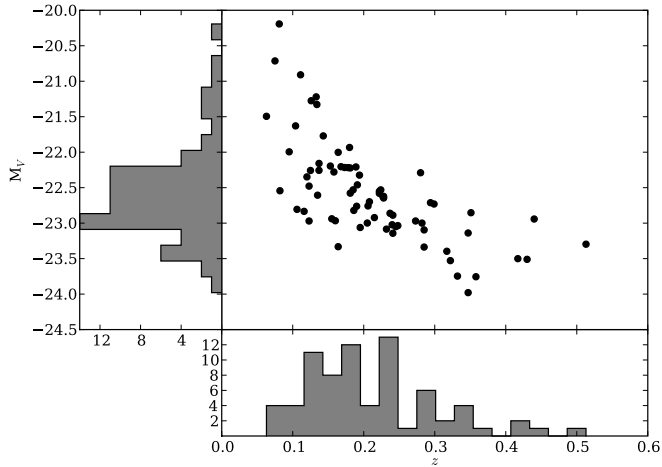}
 \caption{The redshift and magnitude distributions for SLACS lenses. The strong correlation between luminosity and redshift is the result of the selection function of the SLACS lenses, which are drawn from the magnitude-limited SDSS spectroscopic survey.}
 \label{F_redshift_mag}
\end{figure}

The final set of SLACS lenses spans a redshift range from $z = 0.063$ to $z = 0.513$, with a median redshift of $z_{\rm med} = 0.19$. There is a strong correlation between the redshift and the mass of the lensing galaxy (as parameterized by the Einstein radius $r_{\rm Ein}$; see Figure \ref{F_redshift_rein}). This is largely the result of how SLACS lenses are selected; the SDSS spectra constitute a magnitude-limited survey\footnote{Note, however, that the SDSS luminous red galaxy survey has a fainter magnitude limit as well as color criteria.} and the highest redshift objects are therefore more luminous (Figure \ref{F_redshift_mag}) and hence more massive.

\begin{figure}[ht]
 \centering
 \includegraphics[width=0.48\textwidth,clip]{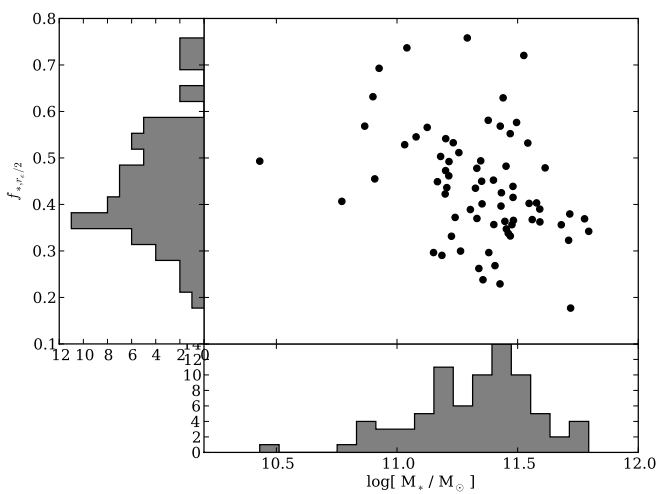}
 \caption{The distributions of the stellar mass, M$_*$, and stellar mass fraction within $r_e/2$, $f_{*,r_{e/2}}$. There is no strong correlation between stellar mass fraction and stellar mass; the relationship between the stellar mass and other observables is explored in more detail in Paper X.}
 \label{F_stellar_masses}
\end{figure}

The 85 SLACS lenses span approximately one decade in stellar and total mass, and are approximately evenly distributed across this range. The mean stellar mass-to-light ratio (M/L) in the V band is 3.2 (5.7) for a Chabrier (Salpeter) IMF; the average total M/L within half of $r_{\rm e}$ ($r_{e/2}$) derived from lensing is 8.7 for either IMF, indicating that some dark matter must be present in both scenarios. We show the distribution of M$_*$ and the stellar mass fraction within $r_{e/2}$, $f_{*,r_{e/2}}$, for the SLACS lenses in Figure \ref{F_stellar_masses}. We use $r_{e/2}$ because it is well matched to the Einstein radii of the lenses and therefore does not depend strongly on the profile of the lens model (e.g., Paper VII). The lack of a strong correlation between these parameters, and correlations between these parameters and the total mass and stellar velocity dispersion, will be explored in more detail in Paper X. The mean stellar mass fraction within the Einstein radius of the lenses is 0.4 (0.7) with a scatter of 0.1 (0.2) for a Chabrier (Salpeter) IMF. In principle, the combination of lensing and dynamics can independently probe this fraction and will therefore discriminate between different IMF models; we leave this analysis for a forthcoming paper (Treu et al. 2009, in preparation).

\begin{figure}[ht]
 \centering
 \includegraphics[width=0.48\textwidth,clip]{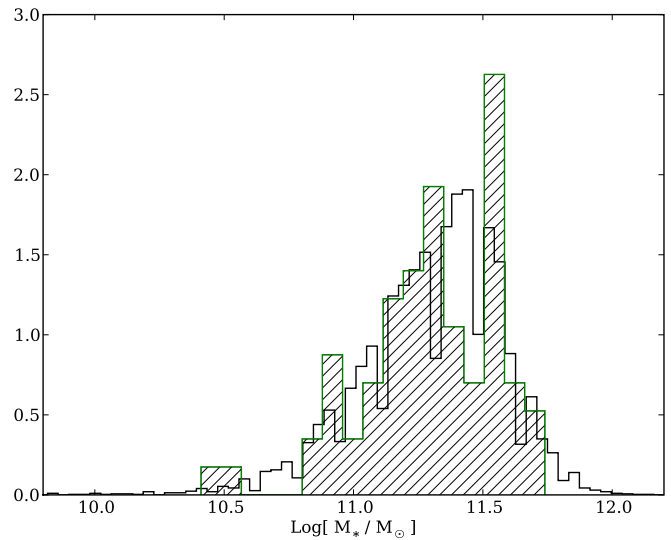}
 \caption{The distributions of stellar masses for the SDSS comparison sample (black, open) and the SLACS lenses (green, hatched). The distributions are indistinguishable (the K-S p-value is $\sim0.95$), strengthening our conclusion that SLACS lenses do not appear to be biased with respect to massive early-type galaxies with similar stellar velocity dispersions.}
 \label{F_sdss_twins}
\end{figure}

As with previous SLACS papers \citep[Paper II, Paper V, Paper VIII; also see][]{auger}, we want to test any biases of the SLACS lenses compared to the parent population of SDSS galaxies from which they are drawn. In the context of this paper, this constitutes a comparison of the stellar masses of SLACS lenses to the stellar masses of SDSS galaxies with similar properties. We use the MPA/JHU SDSS stellar mass catalog to find stellar mass estimates for 100 SDSS galaxies which have a redshift within $\delta z/z < 0.1$ of the lensing galaxy and a velocity dispersion that is within 10\% of the lens velocity dispersion (twelve systems did not have 100 comparison galaxies or did not have masses in the MPA/JHU catalog and are therefore excluded from this analysis). The distributions of stellar masses for the SLACS lenses and for the comparison sample is shown in Figure \ref{F_sdss_twins}; the SLACS lenses have stellar masses which are indistinguishable from their parent population.

It has been suggested that strong gravitational lenses are likely to be more concentrated in mass than non-lensing galaxies \citep[e.g.,][]{vdv,mandelbaum}. This increased concentration may result from triaxiality, where the strong lens is preferentially aligned along the long axis. If this is the case for the SLACS lenses, and if the light follows the same distribution as the total mass, we would expect the SLACS galaxies to have smaller ellipticities than a comparison sample of galaxies (preferential alignment along the long axis will, on average, decrease the ellipticity). Note that the implications of triaxiality also depend on the number of lensed images formed \citep{rozo,mandelbaum} and two-image lenses, four-image lenses, and non-lenses would each yield different distributions of ellipticities if there is a significant bias due to triaxiality. However, Paper VIII has shown that the SLACS lenses are dominated by two-image lenses (approximately 85\% are classified as doubles; SLACS does not appear to have a bias towards finding quad lenses), and we therefore investigate the ellipticities of the full sample of lenses in comparison with non-lenses.

\begin{figure}[ht]
 \centering
 \includegraphics[width=0.48\textwidth,clip]{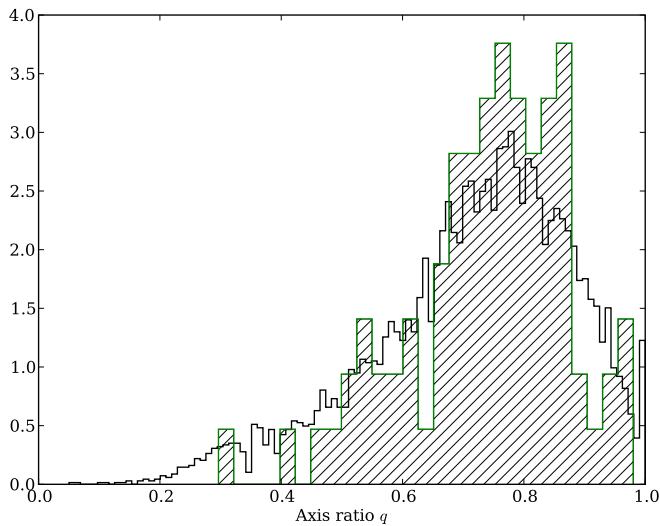}
 \caption{The distributions of axis ratios for the SDSS comparison sample (black, open histogram) and the SLACS lenses (green, hashed histogram). The SLACS lenses do not appear to be more circular than the comparison sample (i.e., the green histogram is not skewed toward 1), indicating a lack of evidence for a preferred axis for the lenses. The two-sample K-S probability is $\sim0.13$ for the two distributions, indicating that the two samples cannot be distinguished at greater than $\sim 1.5\sigma$ confidence.}
 \label{F_ellipticity_comparison}
\end{figure}

We show the distribution of axis ratios $q$ for SLACS lenses (for consistency we use the SDSS axis ratios, but we find that these are very close to the axis ratios inferred from the HST photometry) compared to the distribution of axis ratios for the comparison sample of galaxies in Figure \ref{F_ellipticity_comparison}. We do not find that the SLACS lenses are anomalously circular or elliptical; there is no evidence from the light distribution that SLACS lenses are more concentrated than the comparison sample, although the effect should be quite small and may be unobservable with so few lenses \citep[e.g.,][]{rozo}. A decomposition of the luminous and dark matter components would provide dark matter concentrations that could be compared with simulations, but that test is beyond the scope of this paper.

\section{Discussion and Summary}
We have presented multi-band HST observations of the complete sample of SLACS gravitational lenses. These data have been used to upgrade 6 lenses from Paper V from possible to genuine (grade `A') lens systems as well as confirming 9 newly investigated lens systems. We also present a novel fully-Bayesian stellar mass estimation code and have used this code with our HST photometry to derive stellar masses for each of the grade `A' lens systems. The new results of this work include:

\begin{itemize}
\item  The SLACS program has confirmed 15 new grade `A' lenses. SLACS has now discovered 85 confirmed lenses and has found 13 high-quality lens candidates, yielding a total of 98 likely lens systems. We have created SIE lens models with parametric sources for 11 of the new lenses to allow for accurate determination of the mass within the Einstein radii of the lenses. Approximately 80\% of the lenses have early type morphologies, while $\sim 10\%$ have spiral structure and the other $\sim 10\%$ have S0 morphologies. 

\item  HST photometric colors have been used to determine accurate and precise stellar masses for the SLACS lenses. Typical statistical errors on the mass estimates are 0.1 dex, and the masses determined using HST photometry are equivalent to those estimated from SDSS photometry. We note, however, that G09 find systematically lower masses for the less massive systems ($\sim 0.1$ dex lower at 10$^{11}$ M$_\odot$). Assuming a Chabrier IMF, the mean stellar mass fraction within the Einstein radius of the lenses is 0.4 with a scatter of 0.1; the fraction increases to 0.7 with a scatter of 0.2 for a Salpeter IMF. The total mass-to-light ratio within half of the effective radius is 8.7.

\item There are no substantial differences between the stellar masses and/or ellipticities of SLACS lenses and a comparison sample of SDSS `twin' galaxies. Thus, the SLACS sample appears to be, in effect, a velocity dispersion-selected sample of galaxies with the same properties as similarly-selected comparison samples from SDSS: the SLACS lenses follow the same Fundamental Plane relationship (Paper II; Paper VII), reside in the same environments \citep[Paper VIII;][]{auger}, and have the same distributions of stellar mass and ellipticity as non-lensing galaxies (this work).
\end{itemize}


\acknowledgments 
We thank J. Brinchmann, C. Conroy, C. Grillo, and C. Maraston for useful discussions regarding stellar masses and SPS codes. We also thank C. Nipoti for helpful insights and M. Blanton and D. Hogg for help regarding SDSS photometry. TT acknowledges support from the NSF thorough CAREER award NSF-0642621, by the Sloan Foundation through a Sloan Research Fellowship and by the Packard Foundation through a Packard Fellowship. LK is supported through an NWO-VIDI program subsidy (project number 639.042.505). KB acknowledges support from NASA through Hubble Fellowship grant HST-HF-01215.01-A awarded by the Space Telescope Science Institute (STScI). The work of LAM was carried out at the Jet Propulsion Laboratory, California Institute of Technology, under a contract with NASA. Support for programs \#10494, \#10798, and \#11202 was provided by NASA through grants from the Space Telescope Science Institute. STScI is operated by the Association of Universities for Research in Astronomy, Inc., under NASA contract NAS5-26555. This work has made use of the SDSS database. Funding for the SDSS and SDSS-II has been provided by the Alfred P. Sloan Foundation, the Participating Institutions, the National Science Foundation, the U.S. Department of Energy, the National Aeronautics and Space Administration, the Japanese Monbukagakusho, the Max Planck Society, and the Higher Education Funding Council for England.


\begin{appendix}
\section{Stellar Population Analysis Code}
We present a novel Bayesian stellar population analysis code that allows the full posterior of the composite stellar population (CSP) model parameter space to be efficiently probed under the assumption of arbitrary priors, therefore providing error estimates for all quantities derived from these models (including evolved luminosities and so-called `k-corrections'). Additionally, we provide a routine to perform quick, accurate calculations of the Bayesian evidence using the nested sampling algorithm suggested by \citet{skilling}; this allows different model families (i.e. different choices of IMF) to be compared quantitatively. In this Appendix we describe in detail how the code works, we present the various tests we have used to ensure the stability of the code, and we discuss the effects of choosing different priors or different photometric bands when exploring stellar populations.

\subsection{Exploring the Stellar Population Parameter Space}
Conventional stellar population analyses tend to use a grid or discrete space of CSP models when investigating stellar populations. By cleverly choosing the axes of the grid or employing an {\em a posteriori} weighting scheme, informative (but discrete) priors can be placed on the various CSP parameters; however, it is also possible to inadvertently choose pathological priors. Our code circumvents the discrete priors by employing an interpolation scheme over a grid of CSP models that allows the models to be evaluated at arbitrary points within the bounds of the grid. For example, the analysis of SLACS lenses presented in Section 5 imposed a Gaussian prior on the metallicity of the CSP models informed by the velocity dispersion-metallicity correlation. The modeling is performed in a fully Bayesian framework and the user is therefore required to explicitly define priors for each parameter.
 
The code operates in two stages. First, the user supplies a set of filters and a redshift for the object (future versions will likely allow the redshift to be a free parameter). The code then evaluates a model magnitude in each filter on a grid of (appropriately redshifted) spectra produced by CSP codes. These grids can have arbitrary dimensionality (e.g., any number of parameters among age, metallicity, etc.), although for practical computational purposes grids should generally be restricted to dimensionality $\lesssim6$. A spline interpolation model of this grid of AB magnitudes is generated for each filter, and the set of models are then written to disk. The CSP models are undefined outside of this grid, and therefore care must be taken to make the grid sufficiently wide to allow for any physically interesting values of parameters. Furthermore, although the spline interpolation scheme and the smooth nature of most parameters allows for relatively sparsely sampled grids, the density of points should also be chosen to minimize interpolation errors.

These interpolation models can be evaluated very quickly, allowing data to be efficiently analyzed via Markov chain Monte Carlo methods. The second stage of the code, then, is to define priors for each of the parameters in the model and the stellar mass and to explore the posterior of the parameter space given a catalog of photometric measurements. Note that while many codes use colors (i.e., the difference in magnitude between two filters) as the input data, our code uses the filter magnitudes and it is therefore possible to infer masses from single band observations; note, however, that inferences from single band data are not particularly robust. Future versions of the code will likely add the ability to use fluxes to allow for the input of non-detections.

The exploration of parameter space is undertaken in the usual Bayesian manner. The code uses a likelihood function that assumes Gaussian errors on the photometric magnitudes,
$$
{\rm ln} \mathcal{L}(\vec \theta) = -\frac{1}{2} \sum^{filters}_i \frac{(m_i - model_i(\vec \theta))^2}{\sigma_{m_i}^2} + {\rm ln} (2\pi\sigma_{m_i}^2),
$$
where $\vec \theta$ is the vector of model parameters, $m_i$ is the magnitude in filter $i$, $model_i(\vec \theta)$ is the CSP model magnitude evaluated at $\vec \theta$, and $\sigma^2_{m_i}$ is the observational error on the magnitude. The evaluation of the posterior,
$$
\mathcal{P}(\vec \theta) \propto \mathcal{L}(\vec \theta)\prod^{priors}_i \rm{P}(\vec \theta_i)
$$
where P($\vec \theta$) are the priors, is handled by the Python package PyMC. The amount of time required to probe the parameter space depends on the number of data points and the number of components in the CSP model; a converged chain for a typical model with photometric data in four filters is generally obtained in less than one minute using a single processor on a modern workstation.

\subsection{Comparison with Other Methods}
Our code was originally tested against stellar masses determined by more conventional methods for galaxies in the GOODS-North field \citep{bundy}. We accurately recover the stellar mass estimates (Figure 12) when the same priors as \citet{bundy} are used (i.e., implicit discrete priors on the metallicity and reddening) or when uniform priors bounded by the extrema of the \citet{bundy} priors are used. For these data and priors, the discrete and uniform priors lead to the same most likely value for the stellar mass; however, the uniform priors provide more accurate estimates of parameter uncertainties and covariances. We have also tested our code against the sample of SLACS `twin' comparison galaxies discussed in Section 6 and find that we recover the MPA/JHU mass estimates to within 0.02 dex.

\begin{figure}[ht]
 \begin{center}
 \includegraphics[width=0.75\textwidth]{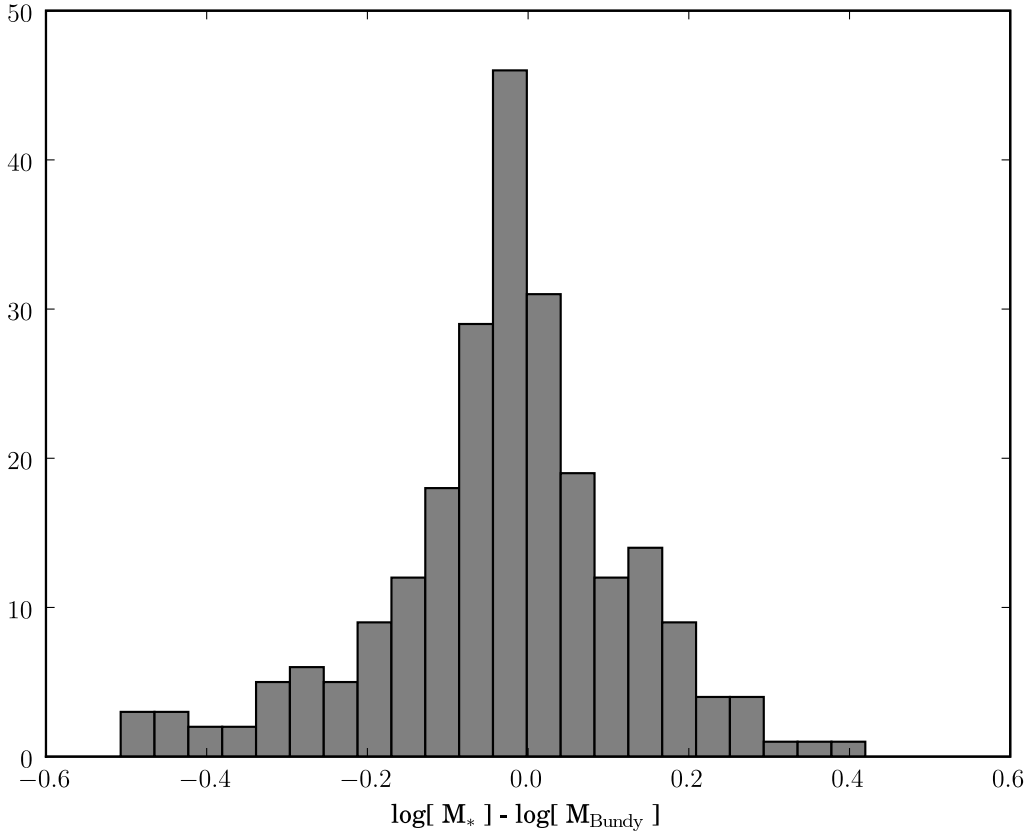}
 \caption{Distribution of offsets in stellar mass estimates for GOODS galaxies as determined by our new stellar mass estimation code and the estimates of \citet{bundy}, assuming a Chabrier IMF. There is a negligible bias, $\Delta$M$ = -0.03$, and the width of the distribution is comparable to the errors on individual mass estimates, $\sigma \approx 0.15$.}
 \end{center}
 \label{F_bundy_comparison}
\end{figure}

\subsection{Effects of Photometric Bands}
The stellar mass of galaxies is most robustly determined using near-infrared bands \citep[e.g.,][but also see Maraston 2005; Conroy et al.\ 2009 for a discussion of potential systematics in the near infrared]{bell}, and we now investigate the effects of increasing the number of filters for a given object. In Figure 13 we show the effects of including the H-band photometry on error estimates for all of the SLACS systems imaged by NICMOS. The red histogram shows the improvement gained when adding H-band data to BVI data, while the blue histogram demonstrates the improvement in precision when adding H-band data to a single optical color (i.e., two optical bands, either B-I or V-I). The inclusion of H-band data tends to decrease the errors by 20-30\%, an improvement that partially results from the posterior of the stellar mass becoming more Gaussian when H-band data are used.

\begin{figure}[ht]
 \begin{center}
 \includegraphics[width=0.75\textwidth]{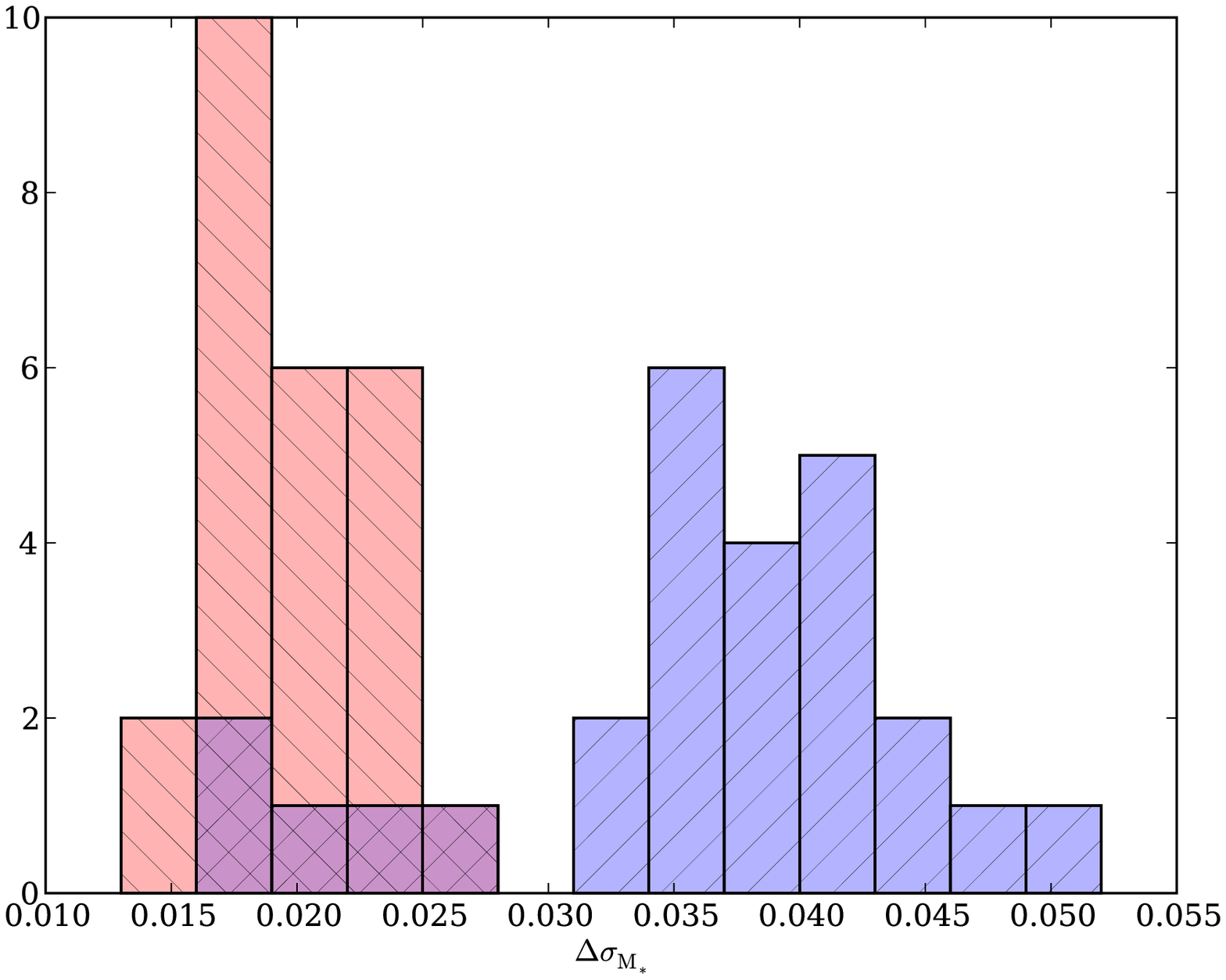}
 \caption{Demonstration of the improvement in errors gained when near infrared bands are used to estimate stellar masses. $\Delta \sigma_{\rm M_*}$ is the difference between $\sigma_{\rm opt}$, the error on the stellar masses of SLACS lenses when only including optical data, and $\sigma_{\rm all}$, the error when the full photometry, including the F160W magnitude, is used. The red histogram is the improvement for lenses with imaging in four bands (three of which are optical) and the blue histogram shows the improvement for lenses with three bands of imaging (two of which are optical); the purple region is the overlap of the two distributions.}
 \end{center}
 \label{F_error_improvement}
\end{figure}

Figures 14 and 15 show the effects of increasing the number of filters on the various parameter degeneracies for one object imaged in four bands, SDSSJ0037-0942. The metallicity constraint with just V and I photometry is weak and tends to reproduce the prior, whereas the addition of H-band data (Figure 14) narrows the inference and slightly shifts the distribution from the peak of the prior. The H-band data also eliminate models with significant dust, although this is not peculiar to the H filter since going from two bands to three bands tends to improve the inference on dust reddening in general. The exclusion of dusty models shifts the peak of the posterior of the stellar mass for this object, decreasing the inferred mass by approximately 0.1 dex. The addition of the B-filter data further constrains the reddening by dust, and the B band also eliminates models with extended star formation histories, again improving the precision of the stellar mass estimate (Figure 15).

\begin{figure}[ht]
 \begin{center}
 \includegraphics[height=0.75\textheight]{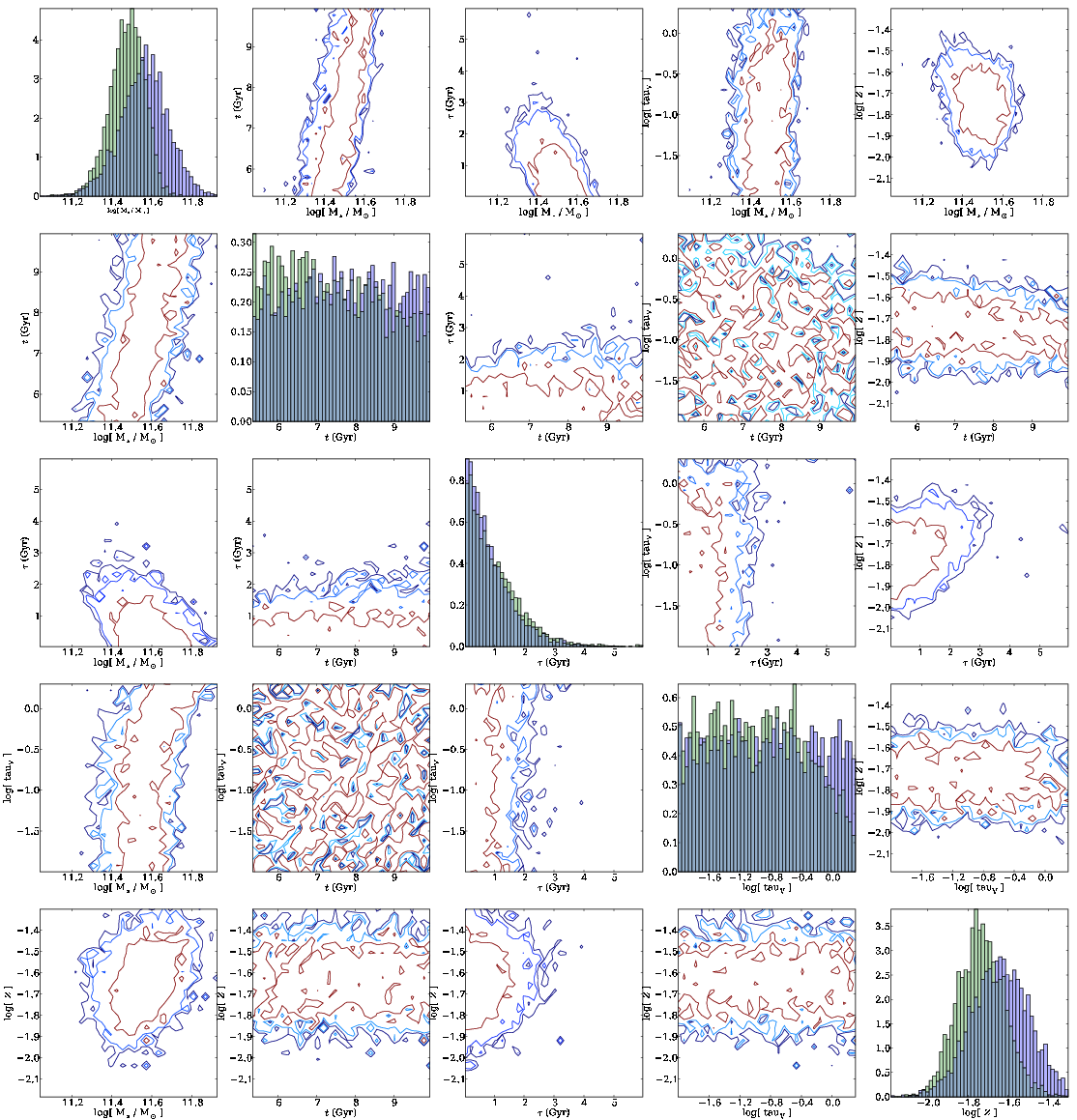}
 \caption{Joint marginalized distributions for the CSP parameters comparing V-I-H photometry (upper-right corner, green on the diagonal) to constraints from V-I photometry (lower-left corner, blue on the diagonal).}
 \end{center}
 \label{F_corner_plotsH}
\end{figure}

\begin{figure*}[ht]
 \begin{center}
 \includegraphics[height=0.75\textheight]{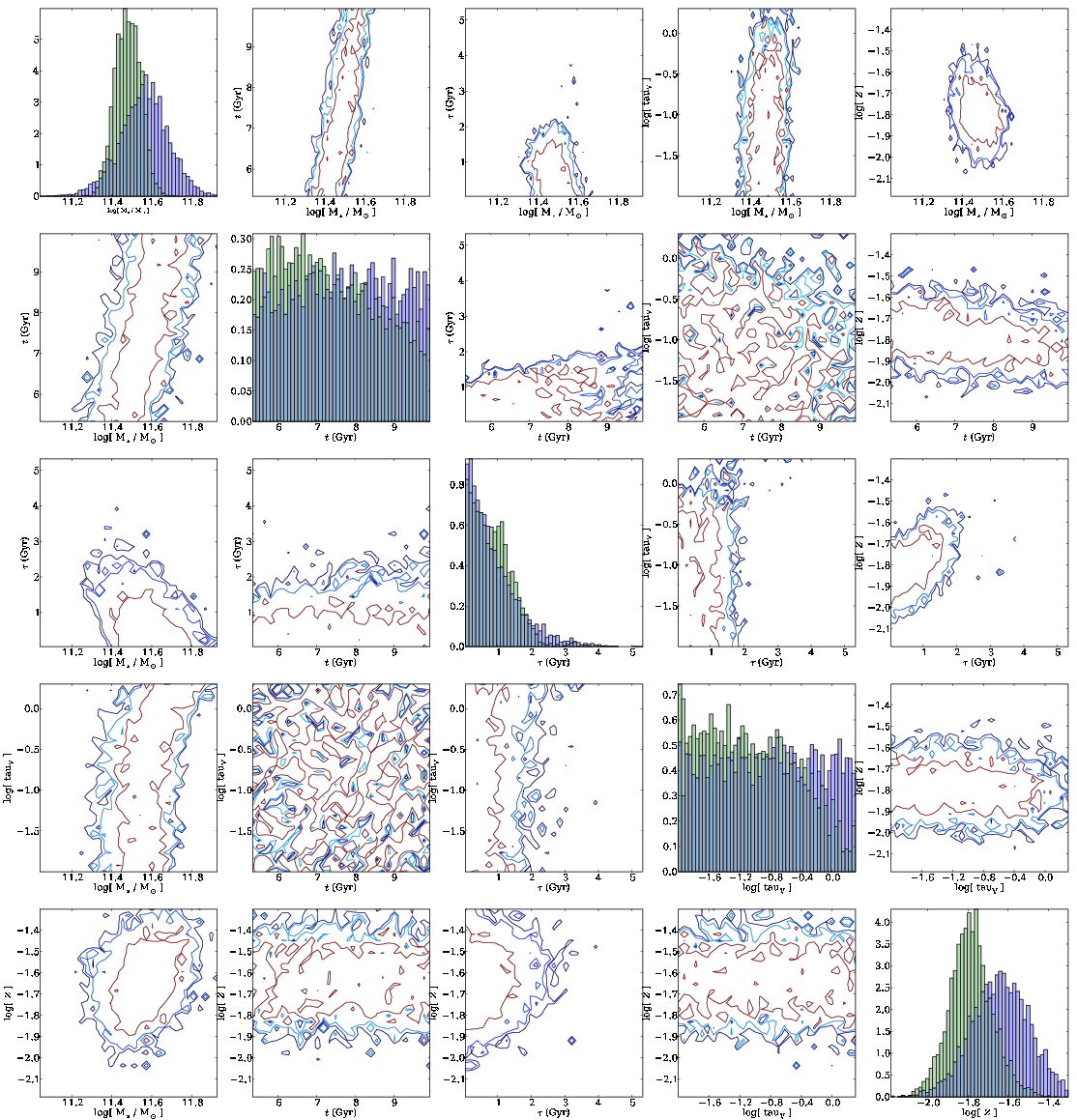}
 \caption{Joint marginalized distributions for the CSP parameters comparing B-V-I-H photometry (upper-right corner, green on the diagonal) to constraints from V-I photometry (lower-left corner, blue on the diagonals); compare with the constraints not using B-band photometry shown in Figure 14.}
 \end{center}
 \label{F_corner_plotsB}
\end{figure*}

\subsection{Effects of Priors}
We have seen in Section 5 that the choice of priors can noticeably affect the inferred stellar mass estimates. We have elected to use informative priors for all of our model parameters except the stellar mass, for which we use a broad uniform prior (note that the uniform prior on the log of $\tau_{\rm V}$ adds more weight to low optical depths, therefore encoding our belief that early-type galaxies are unlikely to be significantly affected by dust reddening). One could use even more constraining priors \citep[by using the correlation between velocity dispersion and $t$ suggested by][for example]{thomas}; however, we now investigate the affects of using less informative priors.

We begin by investigating the prior on the metallicity derived from other SDSS early-type galaxies \citep{gallazzi}. We replace this with a uniform prior on the log of the metallicity, ranging from $Z = 0.0001$ to $Z = 0.05$. Figure 16 shows the effects of the prior on the inferred metallicity; the prior affects the inference for all systems, but most notably for the systems imaged in only two bands (shown as blue diamonds in Figure 16). The corresponding changes in the stellar mass are less dramatic (Figure 17); M$_*$ is robustly determined when three or four photometric bands are used. However, the strong metallicity prior is particularly important for the systems with only two bands of data, as these systems tend to underestimate the stellar mass by $\sim 0.1$ dex without the prior.

\begin{figure}[ht]
 \begin{center}
 \includegraphics[width=0.75\textwidth]{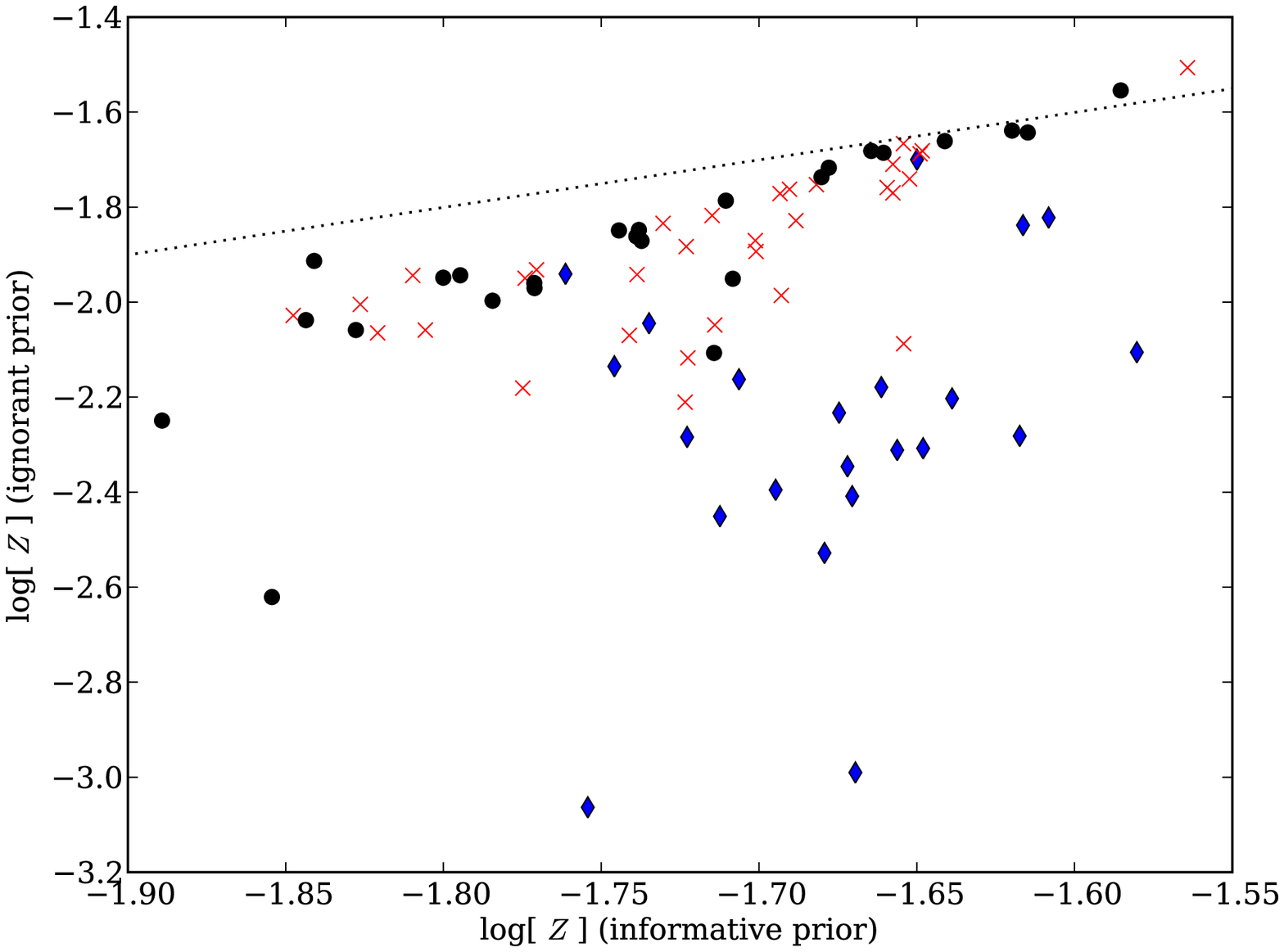}
 \caption{Comparison between the inferred metallicity of SLACS lenses assuming an informative or ignorant prior, where the informative prior is described in Section 5. The dotted line is the identity (note the different axis scales) while the blue diamonds are systems with two bands of imaging, the red crosses were observed in three bands, and the black circles have four bands. Solar metallicity is -1.7.}
 \end{center}
 \label{F_noZ_logZ}
\end{figure}

\begin{figure}[ht]
 \begin{center}
 \includegraphics[width=0.6\textwidth]{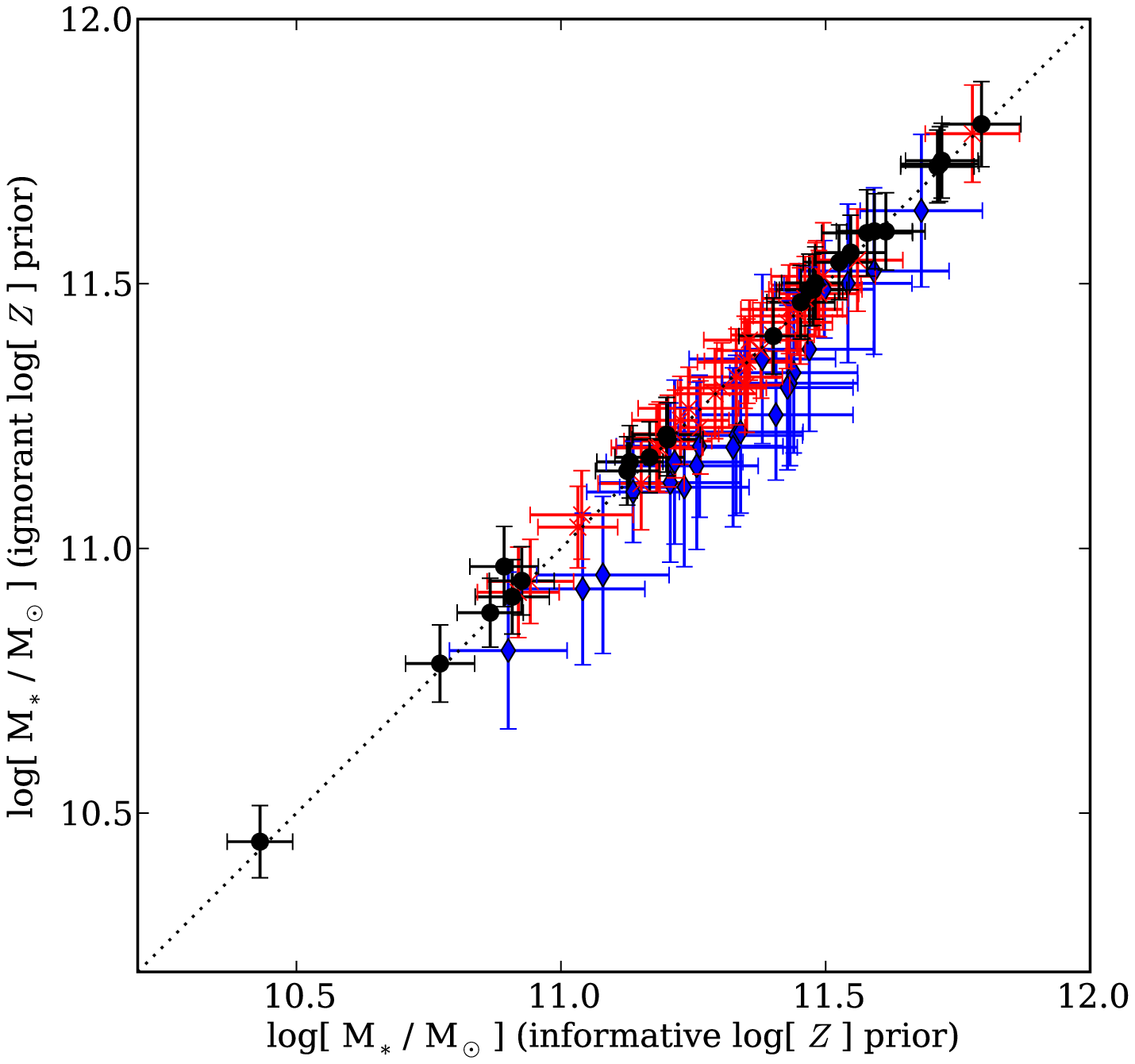}
 \caption{Comparison between the inferred stellar masses of SLACS lenses assuming an informative or ignorant prior on the metallicity. The dotted line is the identity while the blue diamonds are systems with two bands of imaging, the red crosses were observed in three bands, and the black circles have four bands.}
 \end{center}
 \label{F_noZ_mass}
\end{figure}

We also investigate the affects of using `ignorant' priors by loosening the priors on $t$, the time since the beginning of the star formation, and $\tau$, the characteristic time scale of the exponential star formation rate model. We replace these with uniform priors, from 0.6 Gyr to the age of the Universe for $t$ and from 0.04 Gyr to 5.1 Gyr for $\tau$; we also use the uninformed prior on the metallicity described above. The comparison between informative and ignorant priors in shown if Figure 18. The effect of the metallicity prior is evident, but we also find that removing the priors on the age of the stellar population causes the systems observed in only three bands to also shift. The uninformative priors allow for younger stellar populations to be present, thereby underestimating the stellar mass compared to the priors that require somewhat older stellar populations.

\begin{figure}[ht]
 \begin{center}
 \includegraphics[width=0.6\textwidth]{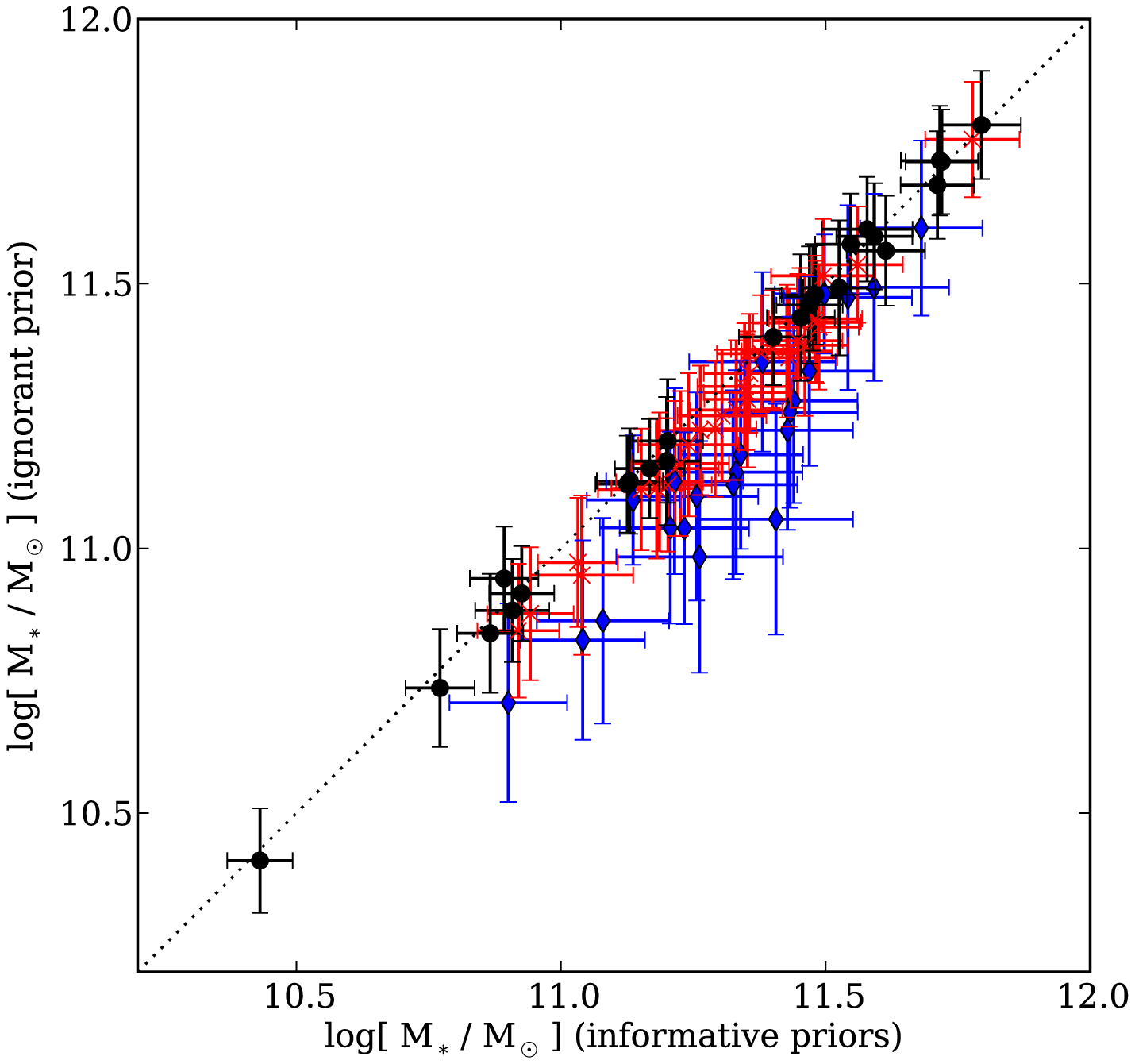}
 \caption{Comparison between the inferred stellar masses of SLACS lenses assuming strong or weak priors on the metallicity and star formation history. The dotted line is the identity while the blue diamonds are systems with two bands of imaging, the red crosses were observed in three bands, and the black circles have four bands.}
 \end{center}
 \label{F_uniform_priors}
\end{figure}

The clear effects of the priors underscore the importance of choosing reasonable priors when inferring stellar masses. Our set of ignorant priors mimics the priors in \citet{bundy}, but we note that those authors investigated a heterogeneous set of galaxies over a range of redshifts, with four bands extending redward only to the observed-frame $z$-band, and the broad priors are therefore appropriate (at least to the extent that using a single set of priors is appropriate). The informed priors used in our analysis of the SLACS lenses are well-motivated, as the SLACS lenses are generally massive early-type galaxies at moderate redshifts ($z \sim 0.2$), which typically have quiescent, older stellar populations.

\end{appendix}

\clearpage
\LongTables
\begin{landscape}
\begin{deluxetable*}{lccccccccccc}[e]
 \tabletypesize{\scriptsize}
 \tablecolumns{12}
 \tablewidth{0pc}
 \tablecaption{Lensing and Stellar Population Properties of SLACS Lenses}
 \tablehead{
   \colhead{Name} &
   \colhead{$r_{\rm Ein}$} &
   \colhead{log[M$_{\rm Ein}$/M$_\odot$]} &
   \colhead{$f^{\rm Chab}_{\rm *,Ein}$} &
   \colhead{$f^{\rm Salp}_{\rm *,Ein}$} &
   \colhead{$\frac{f_{\rm *,Ein}}{f_{\rm *,eff}}$} &
   \colhead{log[M$^{\rm Chab}_*$/M$_\odot$]} &
   \colhead{log[M$^{\rm Salp}_*$/M$_\odot$]} &
   \colhead{$M_{\rm B}$} &
   \colhead{$M_{\rm V}$} &
   \colhead{$M_{\rm B,0}$} &
   \colhead{$M_{\rm V,0}$} \\
   \colhead{} &
   \colhead{(kpc)} &
   \colhead{} &
   \colhead{} &
   \colhead{} &
   \colhead{} &
   \colhead{} &
   \colhead{} &
   \colhead{} &
   \colhead{} &
   \colhead{} &
   \colhead{}
 }
\startdata
SDSSJ0008$-$0004  &     6.59  &    11.55  &   0.27$\pm$0.09  &   0.50$\pm$0.16  &     0.79  &  11.38$\pm$0.14  &  11.64$\pm$0.14  &  -22.25$\pm$0.08  &  -22.94$\pm$0.08  &  -21.55$\pm$0.19 &  -22.33$\pm$0.18 \\
SDSSJ0029$-$0055  &     3.48  &    11.08  &   0.50$\pm$0.14  &   0.89$\pm$0.26  &     0.56  &  11.33$\pm$0.13  &  11.58$\pm$0.13  &  -21.93$\pm$0.12  &  -22.63$\pm$0.07  &  -21.52$\pm$0.12 &  -22.27$\pm$0.12 \\
SDSSJ0037$-$0942  &     4.95  &    11.47  &   0.40$\pm$0.06  &   0.71$\pm$0.10  &     0.78  &  11.48$\pm$0.06  &  11.73$\pm$0.06  &  -22.37$\pm$0.06  &  -23.06$\pm$0.06  &  -22.02$\pm$0.10 &  -22.76$\pm$0.10 \\
SDSSJ0044$+$0113  &     1.72  &    10.96  &   0.37$\pm$0.08  &   0.64$\pm$0.13  &     0.40  &  11.23$\pm$0.09  &  11.47$\pm$0.09  &  -21.67$\pm$0.15  &  -22.35$\pm$0.08  &  -21.42$\pm$0.12 &  -22.14$\pm$0.08 \\
SDSSJ0109$+$1500  &     3.05  &    11.12  &         \nodata  &         \nodata  &  \nodata  &         \nodata  &         \nodata  &          \nodata  &          \nodata  &          \nodata &          \nodata \\
SDSSJ0157$-$0056  &     4.89  &    11.41  &   0.51$\pm$0.12  &   0.90$\pm$0.20  &     0.84  &  11.50$\pm$0.10  &  11.74$\pm$0.10  &  -22.61$\pm$0.07  &  -23.30$\pm$0.06  &  -21.81$\pm$0.22 &  -22.59$\pm$0.20 \\
SDSSJ0216$-$0813  &     5.53  &    11.69  &   0.36$\pm$0.06  &   0.61$\pm$0.10  &     0.56  &  11.79$\pm$0.07  &  12.03$\pm$0.07  &  -23.01$\pm$0.05  &  -23.75$\pm$0.05  &  -22.53$\pm$0.10 &  -23.32$\pm$0.10 \\
SDSSJ0252$+$0039  &     4.40  &    11.25  &   0.40$\pm$0.12  &   0.71$\pm$0.21  &     0.86  &  11.21$\pm$0.13  &  11.46$\pm$0.13  &  -21.57$\pm$0.10  &  -22.29$\pm$0.07  &  -21.12$\pm$0.13 &  -21.90$\pm$0.13 \\
SDSSJ0330$-$0020  &     5.45  &    11.40  &   0.41$\pm$0.08  &   0.69$\pm$0.15  &     0.93  &  11.35$\pm$0.09  &  11.58$\pm$0.09  &  -22.19$\pm$0.09  &  -22.85$\pm$0.06  &  -21.60$\pm$0.16 &  -22.35$\pm$0.16 \\
SDSSJ0405$-$0455  &     1.14  &    10.41  &         \nodata  &         \nodata  &  \nodata  &         \nodata  &         \nodata  &          \nodata  &          \nodata  &          \nodata &          \nodata \\
SDSSJ0728$+$3835  &     4.21  &    11.30  &   0.56$\pm$0.16  &   1.00$\pm$0.29  &     0.82  &  11.44$\pm$0.12  &  11.69$\pm$0.12  &  -22.02$\pm$0.11  &  -22.76$\pm$0.07  &  -21.67$\pm$0.11 &  -22.45$\pm$0.10 \\
SDSSJ0737$+$3216  &     4.66  &    11.46  &   0.41$\pm$0.07  &   0.73$\pm$0.12  &     0.46  &  11.72$\pm$0.07  &  11.96$\pm$0.07  &  -22.78$\pm$0.05  &  -23.53$\pm$0.05  &  -22.32$\pm$0.09 &  -23.11$\pm$0.09 \\
SDSSJ0808$+$4706  &  \nodata  &  \nodata  &         \nodata  &         \nodata  &  \nodata  &  11.26$\pm$0.16  &  11.51$\pm$0.15  &  -22.49$\pm$0.15  &  -23.00$\pm$0.07  &  -21.94$\pm$0.14 &  -22.55$\pm$0.12 \\
SDSSJ0819$+$4534  &     2.73  &    11.04  &   0.32$\pm$0.06  &   0.57$\pm$0.11  &     0.49  &  11.15$\pm$0.08  &  11.40$\pm$0.08  &  -21.67$\pm$0.06  &  -22.32$\pm$0.07  &  -21.26$\pm$0.12 &  -21.98$\pm$0.12 \\
SDSSJ0822$+$2652  &     4.45  &    11.38  &   0.41$\pm$0.12  &   0.73$\pm$0.22  &     0.71  &  11.43$\pm$0.13  &  11.69$\pm$0.13  &  -22.19$\pm$0.11  &  -22.89$\pm$0.07  &  -21.75$\pm$0.12 &  -22.51$\pm$0.13 \\
SDSSJ0841$+$3824  &     2.96  &    11.12  &   0.31$\pm$0.10  &   0.54$\pm$0.18  &     0.32  &  11.41$\pm$0.15  &  11.65$\pm$0.14  &  -22.17$\pm$0.18  &  -22.83$\pm$0.08  &  -21.92$\pm$0.14 &  -22.62$\pm$0.09 \\
SDSSJ0903$+$4116  &     7.23  &    11.66  &   0.35$\pm$0.11  &   0.62$\pm$0.20  &     0.81  &  11.59$\pm$0.14  &  11.84$\pm$0.14  &  -22.83$\pm$0.08  &  -23.51$\pm$0.08  &  -22.10$\pm$0.21 &  -22.88$\pm$0.21 \\
SDSSJ0912$+$0029  &     4.58  &    11.60  &   0.34$\pm$0.05  &   0.60$\pm$0.09  &     0.53  &  11.71$\pm$0.07  &  11.96$\pm$0.07  &  -22.56$\pm$0.06  &  -23.33$\pm$0.06  &  -22.28$\pm$0.09 &  -23.09$\pm$0.09 \\
SDSSJ0935$-$0003  &     4.26  &    11.60  &   0.20$\pm$0.03  &   0.35$\pm$0.05  &     0.31  &  11.72$\pm$0.07  &  11.96$\pm$0.07  &  -23.34$\pm$0.06  &  -23.98$\pm$0.06  &  -22.74$\pm$0.15 &  -23.46$\pm$0.14 \\
SDSSJ0936$+$0913  &     3.45  &    11.17  &   0.57$\pm$0.16  &   1.03$\pm$0.29  &     0.63  &  11.43$\pm$0.12  &  11.68$\pm$0.12  &  -22.04$\pm$0.12  &  -22.76$\pm$0.07  &  -21.70$\pm$0.11 &  -22.47$\pm$0.11 \\
SDSSJ0946$+$1006  &     4.95  &    11.46  &   0.26$\pm$0.07  &   0.46$\pm$0.13  &     0.68  &  11.34$\pm$0.12  &  11.59$\pm$0.12  &  -21.83$\pm$0.11  &  -22.55$\pm$0.07  &  -21.44$\pm$0.11 &  -22.22$\pm$0.11 \\
SDSSJ0955$+$0101  &     1.83  &    10.83  &   0.36$\pm$0.05  &   0.64$\pm$0.09  &     0.82  &  10.77$\pm$0.07  &  11.02$\pm$0.06  &  -20.14$\pm$0.06  &  -20.91$\pm$0.05  &  -19.95$\pm$0.07 &  -20.75$\pm$0.07 \\
SDSSJ0956$+$5100  &     5.05  &    11.57  &   0.35$\pm$0.07  &   0.62$\pm$0.12  &     0.72  &  11.56$\pm$0.09  &  11.81$\pm$0.08  &  -22.42$\pm$0.06  &  -23.14$\pm$0.07  &  -22.02$\pm$0.11 &  -22.80$\pm$0.11 \\
SDSSJ0959$+$4416  &     3.61  &    11.23  &   0.55$\pm$0.15  &   0.99$\pm$0.28  &     0.64  &  11.47$\pm$0.12  &  11.72$\pm$0.12  &  -22.13$\pm$0.11  &  -22.86$\pm$0.07  &  -21.72$\pm$0.11 &  -22.51$\pm$0.11 \\
SDSSJ0959$+$0410  &     2.24  &    10.88  &   0.42$\pm$0.07  &   0.73$\pm$0.11  &     0.78  &  10.91$\pm$0.07  &  11.15$\pm$0.06  &  -20.51$\pm$0.06  &  -21.28$\pm$0.05  &  -20.30$\pm$0.08 &  -21.09$\pm$0.08 \\
SDSSJ1016$+$3859  &     3.13  &    11.17  &   0.48$\pm$0.13  &   0.84$\pm$0.23  &     0.82  &  11.23$\pm$0.12  &  11.48$\pm$0.12  &  -21.47$\pm$0.13  &  -22.20$\pm$0.07  &  -21.17$\pm$0.11 &  -21.95$\pm$0.09 \\
SDSSJ1020$+$1122  &     5.12  &    11.54  &   0.45$\pm$0.13  &   0.81$\pm$0.22  &     0.89  &  11.54$\pm$0.12  &  11.80$\pm$0.12  &  -22.24$\pm$0.09  &  -23.00$\pm$0.07  &  -21.79$\pm$0.12 &  -22.60$\pm$0.12 \\
SDSSJ1023$+$4230  &     4.50  &    11.37  &   0.38$\pm$0.11  &   0.67$\pm$0.19  &     0.85  &  11.33$\pm$0.12  &  11.57$\pm$0.12  &  -21.72$\pm$0.12  &  -22.46$\pm$0.07  &  -21.39$\pm$0.11 &  -22.17$\pm$0.10 \\
SDSSJ1029$+$0420  &     1.92  &    10.78  &   0.69$\pm$0.19  &   1.23$\pm$0.32  &     0.76  &  11.04$\pm$0.12  &  11.29$\pm$0.11  &  -20.87$\pm$0.14  &  -21.63$\pm$0.08  &  -20.68$\pm$0.11 &  -21.47$\pm$0.08 \\
SDSSJ1029$+$6115  &  \nodata  &  \nodata  &         \nodata  &         \nodata  &  \nodata  &  11.49$\pm$0.08  &  11.73$\pm$0.08  &  -21.81$\pm$0.10  &  -22.61$\pm$0.07  &  -21.56$\pm$0.10 &  -22.39$\pm$0.08 \\
SDSSJ1032$+$5322  &     2.44  &    11.05  &   0.42$\pm$0.11  &   0.77$\pm$0.20  &     1.18  &  10.90$\pm$0.11  &  11.16$\pm$0.11  &  -20.44$\pm$0.13  &  -21.22$\pm$0.08  &  -20.21$\pm$0.11 &  -21.02$\pm$0.09 \\
SDSSJ1100$+$5329  &     7.02  &    11.67  &   0.33$\pm$0.05  &   0.58$\pm$0.09  &     0.79  &  11.59$\pm$0.07  &  11.84$\pm$0.07  &  -22.71$\pm$0.06  &  -23.40$\pm$0.06  &  -22.22$\pm$0.14 &  -22.96$\pm$0.13 \\
SDSSJ1103$+$5322  &     2.78  &    10.98  &   0.73$\pm$0.13  &   1.28$\pm$0.23  &     0.71  &  11.29$\pm$0.08  &  11.54$\pm$0.08  &  -21.52$\pm$0.11  &  -22.28$\pm$0.07  &  -21.26$\pm$0.11 &  -22.05$\pm$0.09 \\
SDSSJ1106$+$5228  &     2.17  &    10.96  &   0.54$\pm$0.07  &   0.95$\pm$0.13  &     0.73  &  11.13$\pm$0.06  &  11.37$\pm$0.06  &  -21.27$\pm$0.06  &  -22.00$\pm$0.05  &  -21.09$\pm$0.07 &  -21.84$\pm$0.07 \\
SDSSJ1112$+$0826  &     6.19  &    11.65  &   0.33$\pm$0.07  &   0.59$\pm$0.11  &     0.97  &  11.48$\pm$0.09  &  11.73$\pm$0.08  &  -22.25$\pm$0.10  &  -22.97$\pm$0.07  &  -21.77$\pm$0.14 &  -22.56$\pm$0.13 \\
SDSSJ1134$+$6027  &     2.93  &    11.10  &   0.50$\pm$0.13  &   0.89$\pm$0.24  &     0.69  &  11.26$\pm$0.12  &  11.51$\pm$0.12  &  -21.44$\pm$0.12  &  -22.20$\pm$0.07  &  -21.18$\pm$0.10 &  -21.97$\pm$0.09 \\
SDSSJ1142$+$1001  &     3.52  &    11.22  &   0.39$\pm$0.07  &   0.68$\pm$0.13  &     0.64  &  11.30$\pm$0.08  &  11.55$\pm$0.08  &  -21.89$\pm$0.11  &  -22.58$\pm$0.07  &  -21.50$\pm$0.11 &  -22.25$\pm$0.10 \\
SDSSJ1143$-$0144  &     3.27  &    11.29  &   0.26$\pm$0.05  &   0.46$\pm$0.10  &     0.45  &  11.36$\pm$0.09  &  11.60$\pm$0.09  &  -22.16$\pm$0.13  &  -22.81$\pm$0.07  &  -21.92$\pm$0.10 &  -22.61$\pm$0.08 \\
SDSSJ1153$+$4612  &     3.18  &    11.05  &   0.47$\pm$0.13  &   0.84$\pm$0.25  &     0.87  &  11.08$\pm$0.13  &  11.33$\pm$0.13  &  -21.22$\pm$0.13  &  -21.93$\pm$0.07  &  -20.89$\pm$0.11 &  -21.65$\pm$0.10 \\
SDSSJ1204$+$0358  &     3.68  &    11.24  &   0.40$\pm$0.06  &   0.71$\pm$0.10  &     0.88  &  11.20$\pm$0.07  &  11.45$\pm$0.06  &  -21.22$\pm$0.06  &  -22.00$\pm$0.05  &  -20.97$\pm$0.07 &  -21.78$\pm$0.07 \\
SDSSJ1205$+$4910  &     4.27  &    11.40  &   0.36$\pm$0.05  &   0.63$\pm$0.09  &     0.61  &  11.48$\pm$0.06  &  11.72$\pm$0.06  &  -22.19$\pm$0.05  &  -22.92$\pm$0.05  &  -21.86$\pm$0.09 &  -22.63$\pm$0.08 \\
SDSSJ1213$+$6708  &     3.13  &    11.16  &   0.38$\pm$0.08  &   0.66$\pm$0.14  &     0.62  &  11.24$\pm$0.10  &  11.49$\pm$0.09  &  -21.82$\pm$0.15  &  -22.48$\pm$0.08  &  -21.56$\pm$0.12 &  -22.26$\pm$0.08 \\
SDSSJ1218$+$0830  &     3.47  &    11.21  &   0.41$\pm$0.08  &   0.72$\pm$0.13  &     0.60  &  11.35$\pm$0.08  &  11.59$\pm$0.08  &  -21.90$\pm$0.13  &  -22.61$\pm$0.08  &  -21.65$\pm$0.10 &  -22.39$\pm$0.08 \\
SDSSJ1250$+$0523  &     4.18  &    11.26  &   0.68$\pm$0.11  &   1.20$\pm$0.19  &     0.73  &  11.53$\pm$0.07  &  11.77$\pm$0.07  &  -22.37$\pm$0.06  &  -23.08$\pm$0.06  &  -21.98$\pm$0.12 &  -22.75$\pm$0.11 \\
SDSSJ1250$-$0135  &  \nodata  &  \nodata  &         \nodata  &         \nodata  &  \nodata  &  11.13$\pm$0.06  &  11.37$\pm$0.06  &  -21.58$\pm$0.07  &  -22.22$\pm$0.06  &  -21.36$\pm$0.08 &  -22.04$\pm$0.08 \\
SDSSJ1251$-$0208  &     3.03  &    10.98  &   0.34$\pm$0.07  &   0.59$\pm$0.11  &     0.36  &  11.26$\pm$0.08  &  11.50$\pm$0.08  &  -21.84$\pm$0.06  &  -22.53$\pm$0.07  &  -21.46$\pm$0.12 &  -22.20$\pm$0.11 \\
SDSSJ1259$+$6134  &  \nodata  &  \nodata  &         \nodata  &         \nodata  &  \nodata  &  11.50$\pm$0.09  &  11.75$\pm$0.09  &  -22.11$\pm$0.07  &  -22.87$\pm$0.08  &  -21.74$\pm$0.11 &  -22.54$\pm$0.11 \\
SDSSJ1306$+$0600  &     3.87  &    11.33  &   0.27$\pm$0.05  &   0.47$\pm$0.08  &     0.75  &  11.19$\pm$0.08  &  11.43$\pm$0.08  &  -21.51$\pm$0.12  &  -22.22$\pm$0.07  &  -21.20$\pm$0.11 &  -21.95$\pm$0.09 \\
SDSSJ1313$+$4615  &     4.25  &    11.38  &   0.34$\pm$0.07  &   0.61$\pm$0.11  &     0.77  &  11.33$\pm$0.09  &  11.58$\pm$0.08  &  -21.81$\pm$0.12  &  -22.53$\pm$0.07  &  -21.47$\pm$0.11 &  -22.24$\pm$0.10 \\
SDSSJ1313$+$0506  &  \nodata  &  \nodata  &         \nodata  &         \nodata  &  \nodata  &  10.92$\pm$0.08  &  11.16$\pm$0.08  &  -20.71$\pm$0.12  &  -21.45$\pm$0.07  &  -20.46$\pm$0.10 &  -21.23$\pm$0.08 \\
SDSSJ1318$-$0313  &     6.01  &    11.50  &   0.24$\pm$0.05  &   0.42$\pm$0.08  &     0.57  &  11.43$\pm$0.09  &  11.67$\pm$0.09  &  -22.37$\pm$0.10  &  -23.02$\pm$0.07  &  -21.94$\pm$0.12 &  -22.66$\pm$0.12 \\
SDSSJ1319$+$1504  &  \nodata  &  \nodata  &         \nodata  &         \nodata  &  \nodata  &  10.94$\pm$0.08  &  11.19$\pm$0.08  &  -20.71$\pm$0.12  &  -21.46$\pm$0.07  &  -20.44$\pm$0.10 &  -21.22$\pm$0.09 \\
SDSSJ1330$-$0148  &     1.32  &    10.52  &   0.39$\pm$0.06  &   0.69$\pm$0.10  &     0.97  &  10.43$\pm$0.06  &  10.67$\pm$0.06  &  -19.46$\pm$0.06  &  -20.19$\pm$0.06  &  -19.32$\pm$0.07 &  -20.07$\pm$0.07 \\
SDSSJ1402$+$6321  &     4.53  &    11.46  &   0.40$\pm$0.06  &   0.70$\pm$0.10  &     0.65  &  11.55$\pm$0.07  &  11.79$\pm$0.06  &  -22.24$\pm$0.05  &  -23.00$\pm$0.05  &  -21.93$\pm$0.08 &  -22.72$\pm$0.07 \\
SDSSJ1403$+$0006  &     2.62  &    10.98  &   0.54$\pm$0.10  &   0.94$\pm$0.18  &     0.65  &  11.20$\pm$0.08  &  11.44$\pm$0.08  &  -21.49$\pm$0.11  &  -22.21$\pm$0.07  &  -21.15$\pm$0.11 &  -21.92$\pm$0.10 \\
SDSSJ1416$+$5136  &     6.08  &    11.56  &   0.35$\pm$0.06  &   0.61$\pm$0.11  &     1.01  &  11.40$\pm$0.08  &  11.64$\pm$0.08  &  -22.00$\pm$0.09  &  -22.73$\pm$0.07  &  -21.54$\pm$0.12 &  -22.33$\pm$0.11 \\
SDSSJ1420$+$6019  &     1.26  &    10.59  &   0.69$\pm$0.10  &   1.22$\pm$0.17  &     0.63  &  10.93$\pm$0.06  &  11.17$\pm$0.06  &  -20.79$\pm$0.06  &  -21.49$\pm$0.05  &  -20.67$\pm$0.07 &  -21.39$\pm$0.07 \\
SDSSJ1430$+$4105  &     6.53  &    11.73  &   0.33$\pm$0.09  &   0.59$\pm$0.16  &     0.75  &  11.68$\pm$0.12  &  11.93$\pm$0.11  &  -22.58$\pm$0.09  &  -23.34$\pm$0.07  &  -22.12$\pm$0.12 &  -22.94$\pm$0.12 \\
SDSSJ1432$+$6317  &     2.78  &    11.05  &   0.39$\pm$0.07  &   0.69$\pm$0.13  &     0.30  &  11.46$\pm$0.08  &  11.71$\pm$0.09  &  -22.29$\pm$0.13  &  -22.97$\pm$0.07  &  -22.06$\pm$0.10 &  -22.77$\pm$0.08 \\
SDSSJ1436$+$3640  &  \nodata  &  \nodata  &         \nodata  &         \nodata  &  \nodata  &         \nodata  &         \nodata  &          \nodata  &          \nodata  &          \nodata &          \nodata \\
SDSSJ1436$-$0000  &     4.80  &    11.36  &   0.37$\pm$0.07  &   0.65$\pm$0.13  &     0.60  &  11.45$\pm$0.08  &  11.69$\pm$0.09  &  -22.44$\pm$0.10  &  -23.10$\pm$0.06  &  -21.94$\pm$0.13 &  -22.67$\pm$0.13 \\
SDSSJ1443$+$0304  &     1.93  &    10.78  &   0.50$\pm$0.07  &   0.90$\pm$0.13  &     0.83  &  10.87$\pm$0.06  &  11.12$\pm$0.06  &  -20.60$\pm$0.06  &  -21.33$\pm$0.05  &  -20.37$\pm$0.08 &  -21.13$\pm$0.08 \\
SDSSJ1451$-$0239  &     2.33  &    10.92  &   0.47$\pm$0.07  &   0.80$\pm$0.12  &     0.54  &  11.17$\pm$0.07  &  11.39$\pm$0.06  &  -21.57$\pm$0.06  &  -22.26$\pm$0.05  &  -21.34$\pm$0.08 &  -22.06$\pm$0.08 \\
SDSSJ1525$+$3327  &     6.55  &    11.68  &   0.38$\pm$0.08  &   0.66$\pm$0.13  &     0.60  &  11.78$\pm$0.09  &  12.02$\pm$0.09  &  -23.03$\pm$0.08  &  -23.76$\pm$0.06  &  -22.46$\pm$0.15 &  -23.26$\pm$0.14 \\
SDSSJ1531$-$0105  &     4.71  &    11.43  &   0.37$\pm$0.08  &   0.66$\pm$0.13  &     0.75  &  11.43$\pm$0.09  &  11.68$\pm$0.09  &  -22.31$\pm$0.14  &  -22.97$\pm$0.07  &  -21.98$\pm$0.11 &  -22.69$\pm$0.09 \\
SDSSJ1538$+$5817  &     2.50  &    10.95  &   0.48$\pm$0.08  &   0.84$\pm$0.15  &     0.80  &  11.03$\pm$0.08  &  11.28$\pm$0.08  &  -21.05$\pm$0.11  &  -21.77$\pm$0.07  &  -20.81$\pm$0.10 &  -21.56$\pm$0.08 \\
SDSSJ1614$+$4522  &     2.54  &    10.86  &   0.48$\pm$0.15  &   0.88$\pm$0.25  &     0.44  &  11.21$\pm$0.13  &  11.47$\pm$0.12  &  -21.50$\pm$0.13  &  -22.22$\pm$0.07  &  -21.18$\pm$0.12 &  -21.94$\pm$0.10 \\
SDSSJ1618$+$4353  &  \nodata  &  \nodata  &         \nodata  &         \nodata  &  \nodata  &  11.14$\pm$0.09  &  11.39$\pm$0.09  &  -21.65$\pm$0.07  &  -22.30$\pm$0.09  &  -21.19$\pm$0.14 &  -21.92$\pm$0.15 \\
SDSSJ1621$+$3931  &     4.97  &    11.47  &   0.34$\pm$0.05  &   0.60$\pm$0.09  &     0.70  &  11.45$\pm$0.06  &  11.70$\pm$0.07  &  -22.37$\pm$0.06  &  -23.05$\pm$0.06  &  -21.97$\pm$0.11 &  -22.70$\pm$0.10 \\
SDSSJ1627$-$0053  &     4.18  &    11.36  &   0.46$\pm$0.09  &   0.80$\pm$0.16  &     0.73  &  11.45$\pm$0.09  &  11.70$\pm$0.09  &  -21.93$\pm$0.06  &  -22.70$\pm$0.07  &  -21.61$\pm$0.09 &  -22.41$\pm$0.09 \\
SDSSJ1630$+$4520  &     6.91  &    11.69  &   0.39$\pm$0.07  &   0.69$\pm$0.11  &     0.93  &  11.61$\pm$0.07  &  11.86$\pm$0.07  &  -22.24$\pm$0.05  &  -23.04$\pm$0.05  &  -21.88$\pm$0.08 &  -22.71$\pm$0.08 \\
SDSSJ1636$+$4707  &     3.96  &    11.25  &   0.53$\pm$0.10  &   0.95$\pm$0.18  &     0.78  &  11.38$\pm$0.08  &  11.63$\pm$0.08  &  -21.91$\pm$0.06  &  -22.65$\pm$0.07  &  -21.55$\pm$0.10 &  -22.33$\pm$0.10 \\
SDSSJ1644$+$2625  &     3.07  &    11.12  &   0.46$\pm$0.09  &   0.81$\pm$0.15  &     0.79  &  11.18$\pm$0.09  &  11.43$\pm$0.08  &  -21.45$\pm$0.13  &  -22.16$\pm$0.07  &  -21.19$\pm$0.10 &  -21.94$\pm$0.09 \\
SDSSJ1709$+$2324  &     4.94  &    11.43  &         \nodata  &         \nodata  &  \nodata  &         \nodata  &         \nodata  &          \nodata  &          \nodata  &          \nodata &          \nodata \\
SDSSJ1718$+$6424  &  \nodata  &  \nodata  &         \nodata  &         \nodata  &  \nodata  &  10.89$\pm$0.06  &  11.13$\pm$0.06  &  -21.19$\pm$0.07  &  -21.79$\pm$0.05  &  -20.96$\pm$0.07 &  -21.61$\pm$0.07 \\
SDSSJ1719$+$2939  &     3.89  &    11.28  &   0.40$\pm$0.08  &   0.71$\pm$0.14  &     0.94  &  11.22$\pm$0.08  &  11.46$\pm$0.08  &  -21.50$\pm$0.12  &  -22.22$\pm$0.07  &  -21.17$\pm$0.10 &  -21.94$\pm$0.10 \\
SDSSJ2141$-$0001  &  \nodata  &  \nodata  &         \nodata  &         \nodata  &  \nodata  &         \nodata  &         \nodata  &          \nodata  &          \nodata  &          \nodata &          \nodata \\
SDSSJ2238$-$0754  &     3.08  &    11.11  &   0.41$\pm$0.06  &   0.73$\pm$0.11  &     0.67  &  11.20$\pm$0.06  &  11.45$\pm$0.06  &  -21.55$\pm$0.06  &  -22.26$\pm$0.06  &  -21.31$\pm$0.09 &  -22.05$\pm$0.09 \\
SDSSJ2300$+$0022  &     4.51  &    11.47  &   0.33$\pm$0.05  &   0.58$\pm$0.09  &     0.77  &  11.40$\pm$0.07  &  11.65$\pm$0.07  &  -21.86$\pm$0.05  &  -22.62$\pm$0.05  &  -21.52$\pm$0.08 &  -22.32$\pm$0.08 \\
SDSSJ2302$-$0840  &  \nodata  &  \nodata  &         \nodata  &         \nodata  &  \nodata  &  11.04$\pm$0.10  &  11.27$\pm$0.10  &  -21.32$\pm$0.17  &  -21.97$\pm$0.08  &  -21.12$\pm$0.14 &  -21.80$\pm$0.08 \\
SDSSJ2303$+$1422  &     4.35  &    11.42  &   0.34$\pm$0.05  &   0.59$\pm$0.09  &     0.61  &  11.47$\pm$0.06  &  11.71$\pm$0.06  &  -22.23$\pm$0.06  &  -22.94$\pm$0.06  &  -21.96$\pm$0.09 &  -22.71$\pm$0.09 \\
SDSSJ2321$-$0939  &     2.47  &    11.08  &   0.48$\pm$0.09  &   0.84$\pm$0.15  &     0.52  &  11.35$\pm$0.08  &  11.60$\pm$0.08  &  -21.83$\pm$0.06  &  -22.54$\pm$0.06  &  -21.65$\pm$0.08 &  -22.40$\pm$0.09 \\
SDSSJ2341$+$0000  &     4.50  &    11.35  &   0.38$\pm$0.07  &   0.67$\pm$0.13  &     0.57  &  11.48$\pm$0.08  &  11.73$\pm$0.08  &  -22.08$\pm$0.11  &  -22.82$\pm$0.07  &  -21.76$\pm$0.11 &  -22.55$\pm$0.10 \\
SDSSJ2343$-$0030  &     4.62  &    11.49  &         \nodata  &         \nodata  &  \nodata  &         \nodata  &         \nodata  &          \nodata  &          \nodata  &          \nodata &          \nodata \\
SDSSJ2347$-$0005  &     6.10  &    11.67  &   0.35$\pm$0.07  &   0.62$\pm$0.12  &     0.86  &  11.58$\pm$0.09  &  11.83$\pm$0.08  &  -22.82$\pm$0.06  &  -23.50$\pm$0.06  &  -22.17$\pm$0.18 &  -22.93$\pm$0.16 \\
\vspace{-7pt}
\enddata
\tablecomments{$M_{\rm B}$ and $M_{\rm V}$ are rest-frame Johnson B and V AB magnitudes inferred from our CSP modeling. $M_{\rm B,0}$ and $M_{\rm V,0}$ are the same, except the CSP models have been evolved to $z = 0$ (the ages have been increased by the amount of cosmic time between the redshift of the galaxy and $z = 0$, and the CSP models are re-evaulated with these new ages).}
\label{T_grade_A_model}
\end{deluxetable*}
\clearpage
\end{landscape}

\end{document}